\documentclass[10pt,draftcls,journal,transmag]{IEEEtran}
\pdfoutput=1
\usepackage{amssymb}
\usepackage{bm}
\usepackage{amsmath}
\usepackage{verbatim}
\usepackage{graphicx}
\usepackage{psfrag}
\usepackage{cite}
\usepackage[caption=false,font=footnotesize]{subfig}
\allowdisplaybreaks
\newcommand{\va}{\bm{a}}
\newcommand{\vh}{\bm{h}}
\newcommand{\vi}{\bm{i}}
\newcommand{\vm}{\bm{m}}
\newcommand{\vmc}{\check{\bm{m}}}
\newcommand{\vb}{\bm{b}}
\newcommand{\ve}{\bm{e}}
\newcommand{\vf}{\bm{f}}
\newcommand{\vj}{\bm{j}}
\newcommand{\vn}{\bm{n}}
\newcommand{\vu}{\bm{u}}

\newcommand{\vp}{\bm{p}}
\newcommand{\vr}{\bm{r}}

\newcommand{\vx}{\bm{x}}
\newcommand{\vy}{\bm{y}}

\newcommand{\vnabla}{\bm{\nabla}}
\newcommand{\vcurl}{{\bm{\nabla}}\times}

\newcommand{\beq}{\begin{equation}}
\newcommand{\eeq}{\end{equation}}

\def\epsilon{\varepsilon}

\begin{document}
\title{On the Energy-Based Variational Model\\ for Vector Magnetic Hysteresis}
\author{\IEEEauthorblockN{Leonid Prigozhin\IEEEauthorrefmark{1},
Vladimir Sokolovsky\IEEEauthorrefmark{2},
John W. Barrett\IEEEauthorrefmark{3}, and
Sergey E. Zirka\IEEEauthorrefmark{4}}
\IEEEauthorblockA{\IEEEauthorrefmark{1}J. Blaustein Institutes for Desert Research, Ben-Gurion University of the Negev, Sede Boqer Campus, 84990 Israel}
\IEEEauthorblockA{\IEEEauthorrefmark{2}Physics Department, Ben-Gurion University of the Negev, Beer-Sheva, 84105 Israel}
\IEEEauthorblockA{\IEEEauthorrefmark{3}Department of Mathematics, Imperial College London, London SW7 2AZ, UK}
\IEEEauthorblockA{\IEEEauthorrefmark{4}Department of Physics and Technology,
Dnepropetrovsk National University, Dnepropetrovsk, 49050 Ukraine}
\thanks{Corresponding author: L. Prigozhin (email: leonid@math.bgu.ac.il).}}
\markboth{}
{Shell \MakeLowercase{\textit{et al.}}: On the Energy-Based Model for Vector Magnetic Hysteresis}

\IEEEtitleabstractindextext{%
\begin{abstract}
 We consider the quasi-static magnetic hysteresis model based on a dry-friction like representation of magnetization. The model has a consistent  energy interpretation, is intrinsically vectorial, and ensures a direct calculation of the stored and dissipated energies at any moment in time, and hence not only on the completion of a closed hysteresis loop.  We discuss the variational formulation of this model and derive an efficient numerical scheme, avoiding the usually employed approximation which can be inaccurate in the vectorial case. The parameters of this model for a nonoriented steel are identified using a set of first order reversal curves.  Finally, the model is incorporated as a local constitutive relation into a 2D finite element simulation accounting for both the magnetic hysteresis and the eddy current.
\end{abstract}

}

\maketitle

\section{Introduction}
The hysteretic constitutive relation between the magnetization and the magnetic field in ferromagnets remains one of the main difficulties in
electromagnetic modeling. The Preisach model \cite{Mgoyz}, providing for, probably, the most accurate macroscopic description for ferromagnetic hysteresis at present, is a black-box-type method for
storing, and using for interpolation, a vast amount of experimental data necessary for the
implementation of this model. 
A simpler and very popular Jiles--Atherton model \cite{JA} needs a patch to avoid a nonphysical behaviour \cite{Carp,MZ}; the physical arguments used for the derivation of this model have been criticized in \cite{Zirka}. Both the Preisach and Jiles--Atherton models are scalar and, although there exist numerous vector modifications, these also lack a true physical justification. Furthermore, in a general situation, the use of these models to predict the evolving magnetization does not make computing the accompanying energy loss straightforward (see, e.g., \cite{FrMgoyz}).

In a seminal work \cite{B97}, Bergqvist proposed a new quasi-static magnetic hysteresis model, phenomenological but having a consistent and genuine energy interpretation, intrinsically vectorial, and ensuring a direct calculation of the stored magnetic energy and the dissipated energy at any moment in time, and not only after the completion of a closed hysteresis loop as is usually the case. This model differs significantly from the previous ones but, like the Jiles--Atherton model, regards the pinning of domain walls as the cause of hysteresis and presents, similarly to the Preisach model, the complex hysteretic behavior as a superposition of reactions of simple hysteretic elements, ``pseudoparticles". Later, the Bergqvist model \cite{B97} and models, closely related to it, have been considered in a series of works;
see \cite{B97b,KrahB04,HenrNicHam2006,HenrHam2006,SEH12,FLavHenr2013,HSHG2014,B14} and the references therein.

To make the magnetization update at each time step explicit  Bergqvist \cite{B97} employed an approximation, turning his vectorial energy-based model into a vector play hysteron model. Such an approximation was used also in almost all following works: the only exception that we know is \cite{FLavHenr2013}, where an optimization problem is solved to find the new value of the magnetization. Although in the scalar case this approximation does not introduce any error at all, in the general vectorial case it leads to an error that does not disappear as the time steps (external field increments) tend to zero.

In this work we avoid such an approximation and propose a more efficient numerical method than in \cite{FLavHenr2013}.
We start with the derivation, and a discussion, of a simplified variational hysteresis model in order to clarify its mathematical structure, then make the model more realistic. We identify the parameters of this model for nonoriented electrical steel using a set of experimental first order reversal curves. Finally, we implement the model as a constitutive relation in a finite element simulation taking into account both the quasi-static hysteretic magnetization and the eddy current.

\section{Energy balance and dry-friction like model of magnetization}

The magnetostatic field energy in a magnetic material can be presented as a sum of the empty space energy, depending on the magnetic field $\vh$, and the internal energy determined by the material magnetization $\vm$. The energy density, \beq W=\frac{1}{2}\mu_0h^2+U(\vm),\label{En}\eeq changes as
\beq \dot{W}=\vh\cdot\dot{\vb}-|r\dot{\vm}|,\label{En_t}\eeq
where $\vb=\mu_0(\vh+\vm)$ is the magnetic induction, $\mu_0$ is the permeability of vacuum, $\vh\cdot\dot{\vb}$ is the rate of the magnetic field work, and $|r\dot{\vm}|$ is the rate of dissipation caused by the irreversible movement of the domain walls accompanying the changes in magnetization \cite{B97}. For an isotropic material the ``friction coefficient" $r$ is a positive scalar; otherwise it is a symmetric positive definite matrix.
Here and below the time derivative of $\vu$ is denoted as $\dot{\vu}$ and, if $\vu$ is a vector, $u$ means $|\vu|$.
Equations (\ref{En}) and (\ref{En_t}) yield
$\mu_0\vh\cdot\dot{\vh}+\vnabla U(\vm)\cdot\dot{\vm} =\mu_0\vh\cdot(\dot{\vh}+\dot{\vm})-|r\dot{\vm}|$
or
\beq (\vh-\vf(\vm))\cdot\dot{\vm}=|k\dot{\vm}|,\label{k}\eeq
where $\vf(\vm)=\frac{1}{\mu_0}\vnabla U(\vm)$ and $k=\frac{1}{\mu_0}r$.

Unlike the Jiles--Atherton model, where the magnetization $\vm$ is assumed to be a sum of its reversible and irreversible parts, the Bergqvist model of hysteresis uses a similar representation for the magnetic field; this difference is crucial. The field   $\vh_r=\vf(\vm)$ is called reversible because the magnetic work it delivers is fully converted into internal energy; the remaining field $\vh_i=\vh-\vh_r$ is called irreversible. Equation (\ref{k}) then takes the form \beq \vh_i\cdot\dot{\vm}=|k\dot{\vm}|.\label{ki}\eeq
For an isotropic material,  (\ref{ki}) is satisfied if  the following ``dry-friction" constitutive relation is postulated:
\beq
\begin{array}{l} |\vh_i|\leq k;\\ \mbox{if}\ |\vh_i|< k\ \mbox{then}\ \dot{\vm}={\bm{0}};\\ \mbox{if}
\ \dot{\vm}\neq{\bm{0}}\ \mbox{it has the direction of}\ \vh_i.\end{array} \label{cr}
\eeq
We note that this multivalued relation is similar to the relation between the rate of plastic deformation and stress in an elasto-plastic material with the yield strength $k$.

To obtain a more convenient formulation of (\ref{cr}) we note that $\vh_i\in \widetilde{K}:=\left\{\vu\in \mathbb{R}^3\ :\ |\vu|\leq k\right\}$ and recall the notion of a subdifferential from convex analysis. Let $f:\ \mathbb{R}^n\rightarrow \mathbb{R}\bigcup\{+\infty\}$ be a convex function which may take also the $+\infty$ values. The set \begin{eqnarray*}\partial f(\vx):=\{\vp\in \mathbb{R}^n\ :\ f(\vy)\geq \ f(\vx)+\vp\cdot(\vy-\vx)\\ \mbox{for all}\ \vy\in \mathbb{R}^n\}\end{eqnarray*} is called the subdifferential of $f$ at the point $\vx$; its elements $\vp\in \partial f(\vx)$ are subgradients of $f$ at $\vx$. If $f$ is differentiable at $\vx$ then
$\partial f(\vx)=\{\nabla f(\vx)\}$ and, if $f(\vx)=+\infty$,  $\partial f(\vx)$ is an empty set.
In addition, if  $\bm{0}\in \partial f(\vx)$ then $f(\vx)\leq f(\vy)$ for all $\vy$.

It is not difficult to find the subdifferential of the indicator function of the set $\widetilde{K}$,
$$I_{\widetilde{K}}(\vx)=\left\{\begin{array}{lr}0&\vx{\in}\widetilde{K},\\ \infty&  \vx{\not \in} \widetilde{K}.\end{array}\right.$$
For $\vx\in \widetilde{K}$ we obtain $\vp\in \partial I_{\widetilde{K}}(\vx)$ if $\vp\cdot(\vy-\vx)\leq 0$ for any $\vy\in \widetilde{K}$. Clearly, if $|\vx|<k$ this condition holds only for $\vp=\bm{0}$ and, if $|\vx|=k$, $\vp$ can be any vector of the same direction as $\vx$. Hence, the
conditions in (\ref{cr}) can be written as \beq \dot{\vm}\in \partial I_{\widetilde{K}}(\vh_i).\label{isoI}\eeq
It follows from the definition of a subdifferential that while $\vh_i$ belongs to the interior of the set $\widetilde{K}$, i.e. $|\vh_i|<k$, the magnetization does not change: $\dot{\vm}=\bm{0}$. Whereas, if $|\vh_i|=k$ then (\ref{isoI}) determines the unique direction of $\dot{\vm}$, since $\partial I_{\widetilde{K}}(\vh_i)=\{\vu\in \mathbb{R}^3\ :\ \vu=\lambda\vh_i,\ \lambda\geq 0\}$.

Until now, the ``dry friction law" was not defined precisely. Now we explain our choice of (\ref{cr}), which does not follow from (\ref{ki}) since it is not the only constitutive relation for which (\ref{ki}) holds. According to a general definition by Moreau (\cite{Moreau}, p. 64), to set a dry friction relation between the irreversible field $\vh_i$ (the ``friction force") and the magnetization velocity $\dot{\vm}$, it is required to define a closed convex set of admissible irreversible fields, $\widetilde{K}$, and postulate the maximal dissipation principle: for a given $\dot{\vm}$ the field $\vh_i$ should
maximize the dissipation power $\mu_0\vh_i\cdot\dot{\vm}$ in the set $\widetilde{K}$. Such a relation between $\dot{\vm}$ and $\vh_i$ is equivalent to (\ref{isoI}), which is equivalent to (\ref{cr}).

In the anisotropic case
we also postulate  that $\dot{\vm}\in \partial I_{\widetilde{K}}(\vh_i)$, where now,  since $k$ is a symmetric positive definite matrix, $\widetilde{K}:=\left\{\vu\in \mathbb{R}^3\ :\ |k^{-1}\vu|\leq 1\right\}.$
In this case, rewriting (\ref{ki}) as $k^{-1}\vh_i\cdot k\dot{\vm}=|k\dot{\vm}|$ we see that this equality holds, since
$\dot{\vm}\in \partial I_{\widetilde{K}}(\vh_i)$ means that
\beq \begin{array}{c}\vh_i\in \widetilde{K}\ \mbox{is such that}\\  \dot{\vm}\cdot(\vu-\vh_i)\leq 0\ \mbox{for any}\  \vu\in\widetilde{K},\end{array}\label{in0} \eeq
which is equivalent to the multivalued constitutive relation
$$\begin{array}{l}|k^{-1}\vh_i|\leq 1;\\ \mbox{if}\ |k^{-1}\vh_i|< 1\ \mbox{then}\ \dot{\vm}=\bm{0};\\ \mbox{if}\ \dot{\vm}\neq \bm{0}\ \mbox{then}\ k\dot{\vm}\ \mbox{has the direction of}\ k^{-1}\vh_i.\end{array}$$
Note that if  $\vh_i\in\widetilde{K}$ then the reversible field $\vh_r=\vh(t)-\vh_i$ belongs to the set $$K(t):=\{\vu\in\mathbb{R}^3\ :\ |k^{-1}(\vh(t)-\vu)|\leq 1\}$$ and the inequality (\ref{in0}) can be rewritten for $\vh_r$:
\beq \begin{array}{c}\vh_r\in K(t)\ \mbox{is such that}\\ \dot{\vm}\cdot(\vu-\vh_r)\geq 0\ \mbox{for any}\ \vu\in K(t).\end{array} \label{in}\eeq
Inverting the dependence $\vh_r=\vf(\vm)$ we obtain that $\vm=\vf^{-1}(\vh_r)$ and, as in \cite{B97}, assume further that the vectors $\vh_r$ and $\vm$ are parallel, i.e.
$\vm=M_{an}(h_r)\frac{\vh_r}{h_r},$ where the anhysteretic function $M_{an}$ 
is non-decreasing and $M_{an}(0)=0$.

Let $S(\vu)=\int_0^uM_{an}(s)ds$. Then
\beq\vm=\vnabla S(\vh_r).\label{nS}\eeq
To solve (\ref{in})--(\ref{nS}) numerically, we substitute (\ref{nS}) into the discretized version of (\ref{in}), $$\begin{array}{c}\vh_r\in K(t)\ \mbox{is such that}\\  (\vm-\check{\vm})\cdot(\vu-\vh_r)\geq 0\ \  \mbox{for any}\  \vu\in K(t),\end{array}$$
where ``$\check{\ }$" means the value from the previous time level. This yields  the variational inequality
\beq\begin{array}{c} \mbox{find}\ \vh_r\in K(t)\ \mbox{such that}\\    (\vnabla S(\vh_r)-\check{\vm})\cdot(\vu-\vh_r)\geq 0\\ \mbox{for any}\  \vu\in K(t),\end{array}\label{vi_hr}\eeq
which is equivalent to an optimization problem:  $\vh_r(t)$ is a solution of
\beq \min_{\vu\in  K(t)}\{S(\vu)-\check{\vm}\cdot\vu\}.\label{opt}\eeq
It can be shown that if the derivative $M'_{an}>0$ then $S$ is a strictly convex function and, since the set $K(t)$ is convex, (\ref{opt}) has a unique solution.

The unconstrained minimum of $S(\vu)-\check{\vm}\cdot\vu$ is at a point $\vu$ where $\vnabla S(\vu)-\check{\vm}=0$;
in this case $\vu=\check{\vh}_r$. Hence, if $|k^{-1}(\vh(t)-\check{\vh}_r)|\leq 1$,  this is a solution also to the constrained problem (\ref{opt}) and $\vm=\check{\vm}$. Otherwise, the equality constraint
$|k^{-1}(\vh(t)-\vh_r(t))|= 1$ holds. We use this observation to solve the optimization problem (\ref{opt}) numerically as follows.

At each time level, the solution to (\ref{opt}) is $\vh_r(t)=\check{\vh}_r$ if $|k^{-1}(\vh(t)-\check{\vh}_r)|\leq 1$. In 2d problems, if this inequality is not true, $\vh_r=\vh(t)+k\vi_{\phi}$,
where $\vi_{\phi}=(\cos\phi,\,\sin\phi)$ is a unit vector and, therefore, one is required to solve an unconstrained 1d minimization problem
\beq \min_{\phi}\{S(\vh(t)+k\vi_{\phi})-\check{\vm}\cdot(\vh(t)+k\vi_{\phi})\}.\label{2dopt}\eeq
Although there can be several local minima, a good initial approximation to the optimal direction $\phi$ is the direction of the vector $k^{-1}(\check{\vh}_r-\vh(t))$. Starting from this approximation, we solved the problem $g'(\phi)=0$, where $g(\phi)=S(\vh(t)+k\vi_{\phi})-\check{\vm}\cdot(\vh(t)+k\vi_{\phi})$, efficiently using Newton's method (see Appendix A). Usually, two or three iterations of this method have been sufficient to find the solution with high accuracy. Although we have only solved 2d problems, this method should be efficient also in 3d problems, where the optimal direction is determined by two angles.

We note that in \cite{B97,B97b,KrahB04,HenrNicHam2006,HenrHam2006,SEH12,HSHG2014,B14} the vector
\beq \vh_r=\vh+k\frac{\check{\vh}_r-\vh(t)}{|\check{\vh}_r-\vh(t)|}\label{Appr1}\eeq
is chosen (in the isotropic case) as the  new value of $\vh_r$ if $|\vh(t)-\check{\vh}_r|> k$. This is equivalent to using our initial approximation for $\phi$ without any further correction and turns (\ref{in})--(\ref{nS}) into a vector play model. In the vectorial case such an approach can introduce an error that does not disappear as the increments of $\vh$ tend to zero (see below).

The hysteresis model (\ref{in})--(\ref{nS}) is oversimplified but it will be used as a building block for a more realistic model (Section III). First, it seems instructive to illustrate the behaviour of this model by several examples. Let us assume, as in \cite{B97}, that
\beq M_{an}(h_r)=\frac{2 m_s}{\pi}\arctan\left(\frac{h_r}{A}\right),\label{Man}\eeq where $m_s$ is the saturation magnetization and the parameter $A$ determines the steepness of the curve. Another popular representation of the anhysteretic curve (see, e.g., \cite{HenrNicHam2006,HenrHam2006,dARTV}) is the Langevin function $M_{an}(h_r)=m_s[\coth(h_r/B)-B/h_r]$, which is very well approximated by (\ref{Man}) if $A=1.7B$. In general, the curve can be approximated by a spline (and  we will use a spline representation of $M_{an}$ to model nonoriented electrical steel in Section \ref{Sec4}). We found that for $m_s=1.23\cdot 10^6$ A/m, $A=38$ A/m and $k=71$ A/m  the model (\ref{in})--(\ref{nS}) with (\ref{Man}) describes well the major hysteresis loop shown in figure 5 of \cite{B14} (here $k$ determines the loop width which is almost constant except close to saturation, where it quickly drops to zero).

Let $\vm(0)=\bm{0}$. First, we set $\vh=(H_m\sin t,0)$. This example is one-dimensional, the approximation (\ref{Appr1}) does not introduce any error and, furthermore, the problem (\ref{2dopt}) can be solved analytically. We used it to check our optimization procedure. Here, and throughout this section and the next, the time step is chosen sufficiently small, about
200--400 time steps per cycle, so that the shown figures are independent of the time step.

\begin{figure}[!h]
\centering
\includegraphics[width=8cm,height=7.3cm]{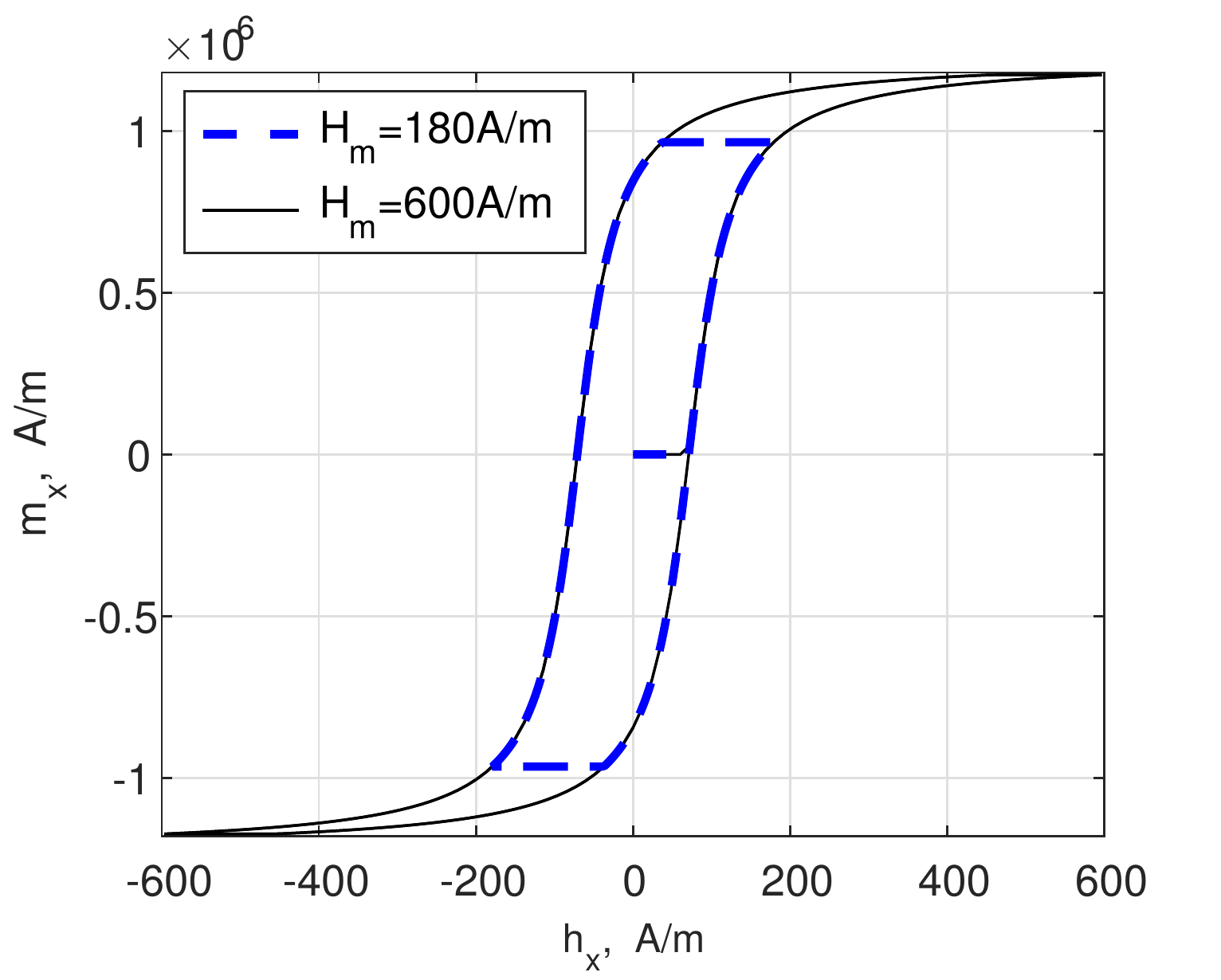}\caption{Two hysteresis loops, simplified model; $\vh=(H_m\sin t,0)$.\label{Fig1}}
\end{figure}

The simulation results for two values of the amplitude $H_m$ (Fig. \ref{Fig1}) show that, although the model's prediction of the major hysteresis loop is correct, the minor loop and the initial magnetization curve are unrealistic.

\begin{figure}[!h]
\centering
\includegraphics[width=8cm]{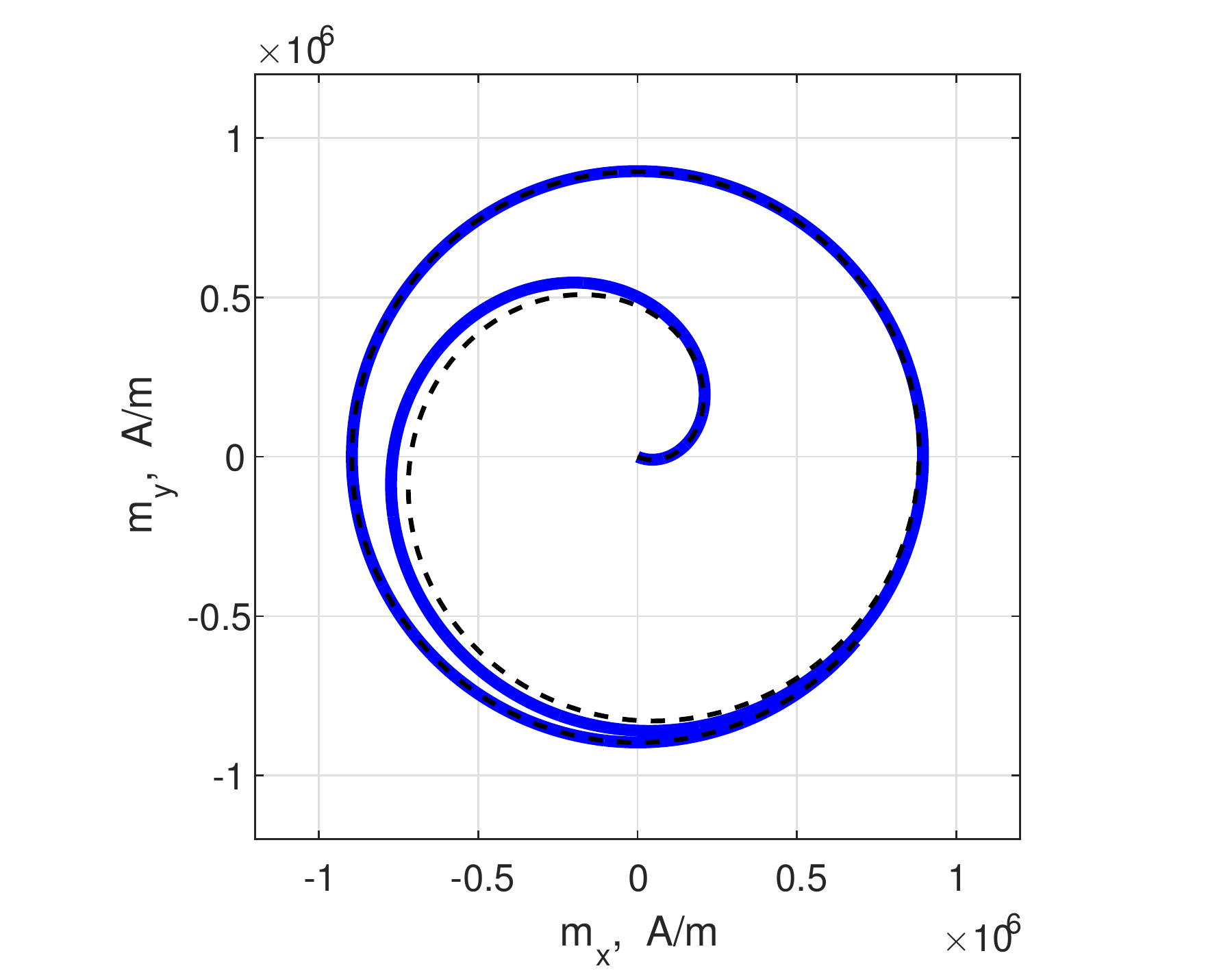}
\caption{Model solution $\vm=(m_x,m_y)$ (solid line) and the explicit approximation (\ref{Appr1}) (dashed line);
  $\vh=H_m(t)(\cos t,\sin t)$, where $H_m(t)=110\min(t/6\pi,1)$ A/m,  $k=71$ A/m.}\label{Fig2}
\end{figure}

\begin{figure}[!h]
\centering
\includegraphics[width=8cm]{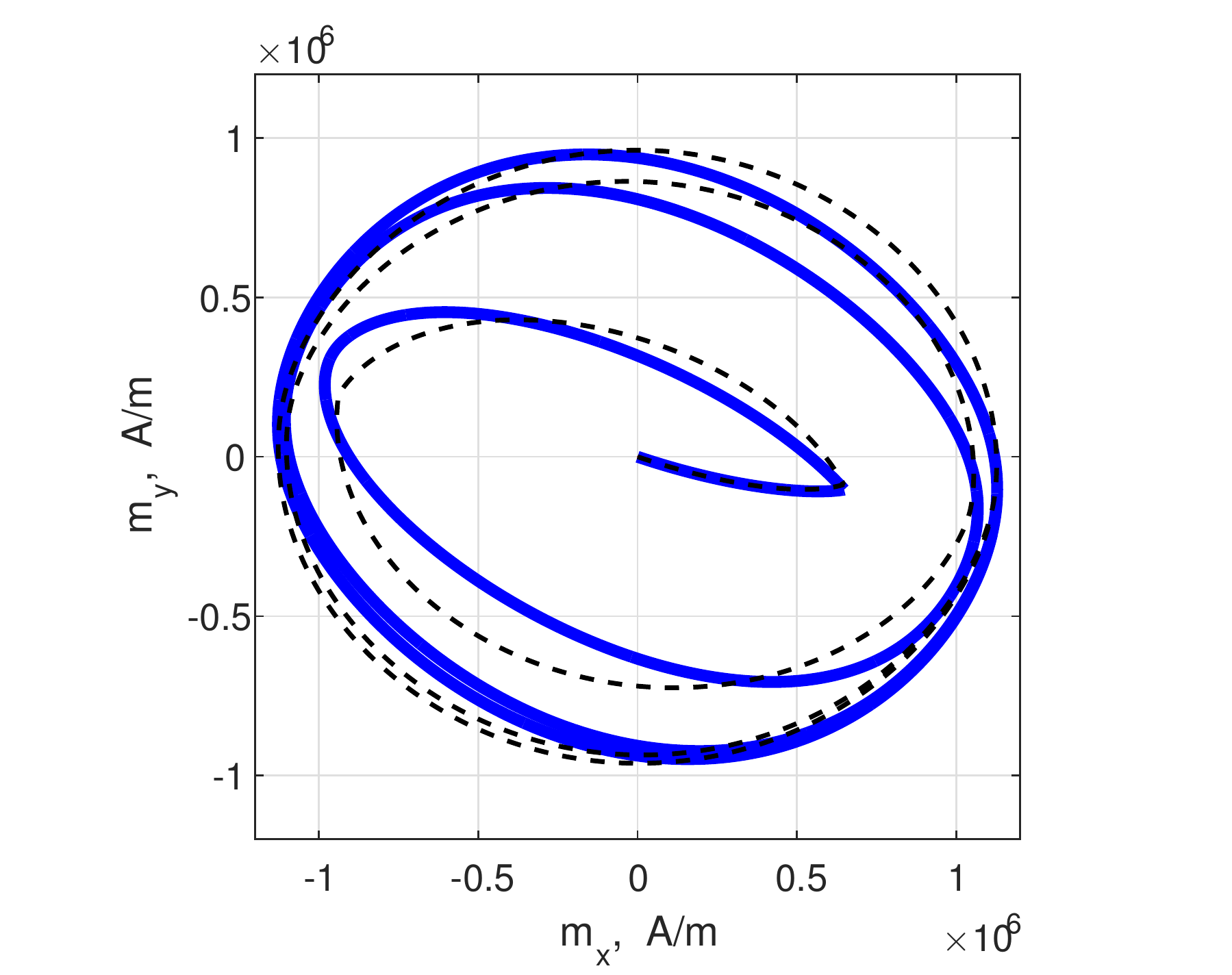}\\ \includegraphics[width=8cm]{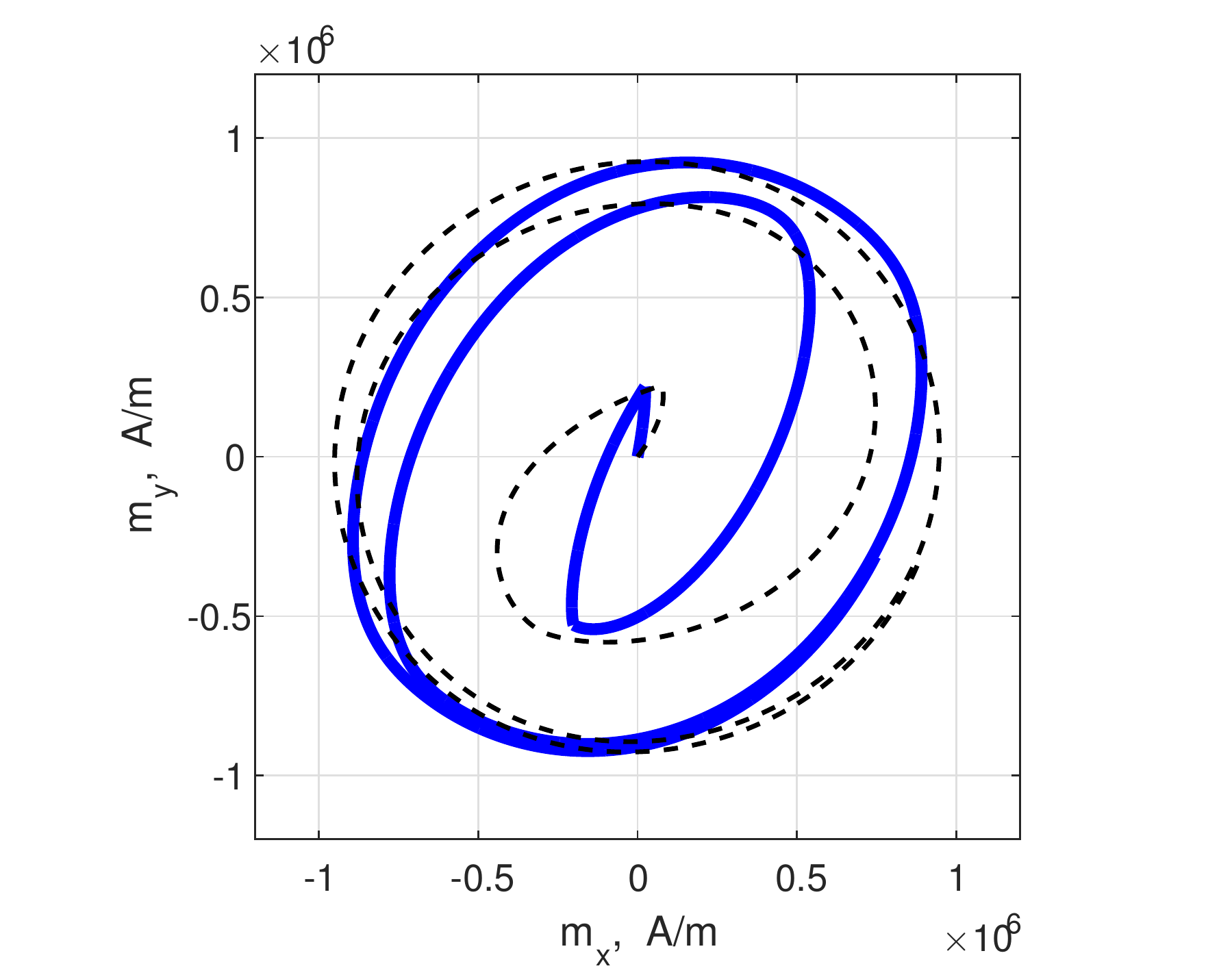}
\caption{Similar to Fig. \ref{Fig2} except: top -- for  $\vh=H_m(t)(3\cos t,\sin t)$; bottom -- for an anisotropic material characterised by the diagonal matrix $k=\mbox{diag}(71,35.5)$ A/m.}\label{Fig3}
\end{figure}

In our next example (Fig. \ref{Fig2}) we assume the magnetic field rotates, $\vh=H_m(t)(\cos t,\sin t)$, with the amplitude $H_m(t)=110\min(t/6\pi,1)$ growing with time until its maximal value 110 A/m is reached. This is a non-scalar situation and we compare the accurate numerical solution of (\ref{2dopt}), equivalent to the time discretized version of (\ref{in}), to the explicit, at each time step, discretized vector play model based on the approximation (\ref{Appr1}). Although the solutions are different in the transient regime, the difference is small and disappears soon after the amplitude of the rotating magnetic field becomes constant. However, the approximation (\ref{Appr1}) is less accurate if the amplitudes of the magnetic field components $h_x$ and $h_y$ are different (Fig. \ref{Fig3}, top) or the material is anisotropic (Fig. \ref{Fig3}, bottom).

\section{A more realistic composite model of a ferromagnetic material}

Prior to using the described energy-based dry-friction like model for modeling hysteresis in real ferromagnets,
this model should be made more realistic. The main modifications, at least partially implemented in all works where such a model has been used,  have been suggested already in \cite{B97,B97b}; their analogues can be found also in some previous models of hysteresis.

First, instead of a single value of the ``friction coefficient" $r$, the material can be characterized by a distribution of $r$ values with the volume density $\omega(r)$; this approach corresponds better to the statistical distribution of the pinning center strengths in the ferromagnetic microstructure. The total magnetization is $\vm=\int \vm^{r}\omega(r)dr$, where each moment $\vm^{r}(t)$ obeys the dry friction model with  its own value of $r$. This can improve the description of the initial magnetization curve and the minor loops.

\begin{figure}[h!]
\begin{center}\includegraphics[width=7cm,height=6cm]{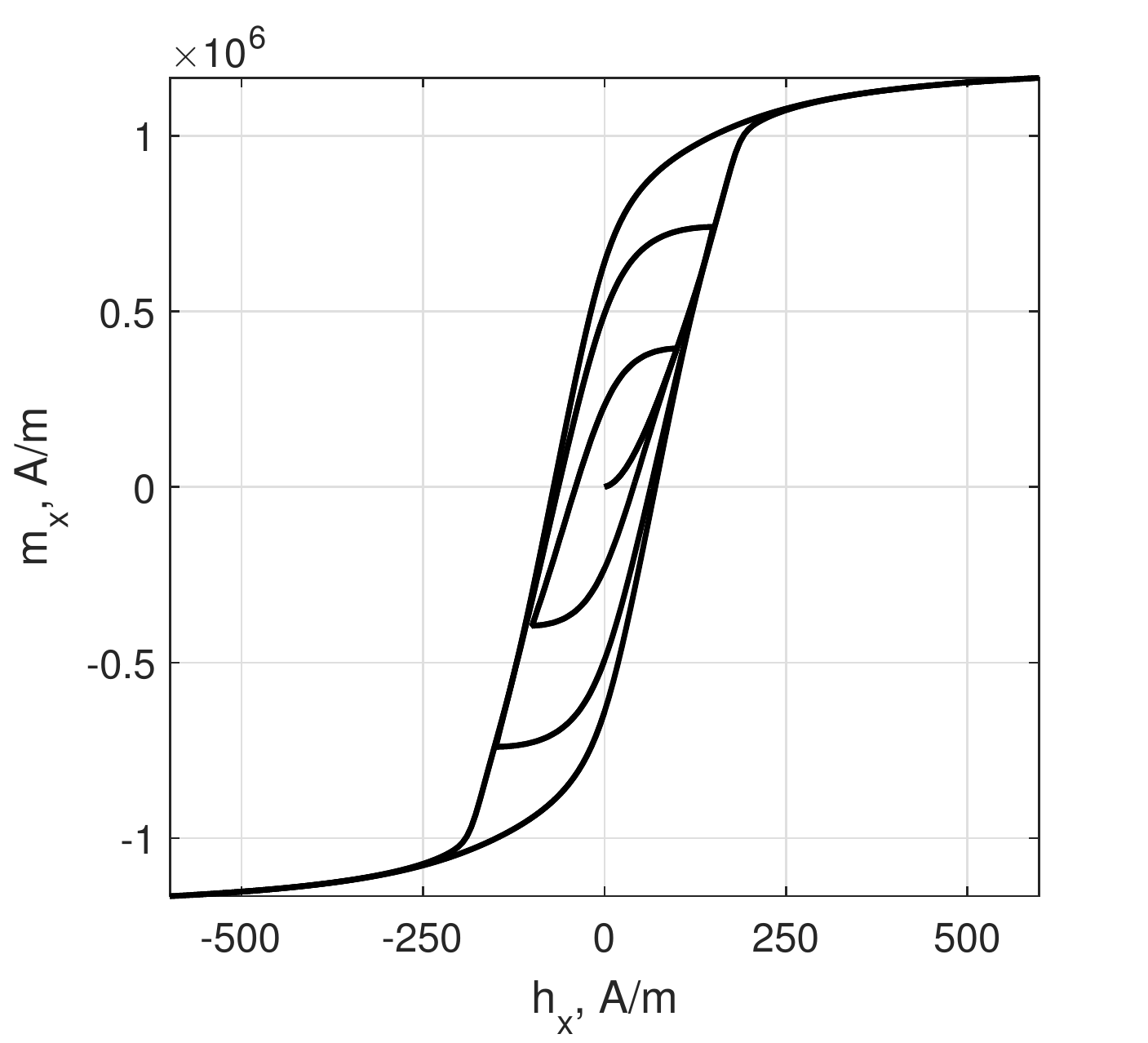}
\caption{Simulation using the composite model.}\label{Fig4}
\end{center}
\end{figure}

For numerical simulations we approximate the distribution by a mixture of $N$ types of pseudoparticles with volume fractions $\omega^l>0$, satisfying $\sum_{l=1}^N\omega^l=1$. Each type is characterised by its own $r=r^l$ and, to account for partial reversibility of the material response \cite{B97,SEH12}  we assign $r=0$ to one of the pseudoparticle types. Overall, we assume
$$\begin{array}{c}W=\frac{1}{2}\mu_0h^2+\sum_{l=1}^N \omega^lU(\vm^l),\\ \vb=\mu_0(\vh+\sum_{l=1}^N \omega^l\vm^l),\\ \dot{W}=\vh\cdot\dot{\vb}-\sum_{l=1}^N \omega^l|r^l\dot{\vm}^l|\end{array}$$
and arrive at an analogue of (\ref{k}),
\beq \sum_{l=1}^N\omega^l\left\{(\vh-\vf(\vm^l))\cdot \dot{\vm}^l-|k^l\dot{\vm}^l|\right\}=0,
\label{kj}\eeq
where $k^l=\frac{1}{\mu_0}r^l$ and, as before, $\vf=\frac{1}{\mu_0}\vnabla U$. We set $\vh_r^l=\vf(\vm^l)$, $\vh_i^l=\vh-\vh_r^l$ and, similarly to what was done above,  satisfy (\ref{kj}) by postulating the constitutive relations
$$\dot{\vm}^l\in \partial I_{\widetilde{K}^l}(\vh_i^l),$$ where $$\widetilde{K}^l:=\{\vu
\in {\mathbb R}^3
\ : \ |(k^l)^{-1}\vu |\leq 1\};$$ if $k^l=0$ we assume $\widetilde{K}^l:=\{\bm{0}\}$. As before, we reformulate these conditions as variational inequalities, similar to (\ref{in})--(\ref{nS}). After discretization in time these inequalities become equivalent to optimization problems similar to (\ref{opt}): we find $\vh_r^l$ on a new time level as a solution to
\beq \min_{\vu\in K^l(t)}\{S(\vu)-\check{\vm}^l\cdot\vu\},\label{optj}\eeq
where $$K^l(t):=\{\vu \in {\mathbb R}^3\ :\ |(k^l)^{-1}(\vh(t)-\vu)|\leq 1\}$$ except for $k^l=0$: in that case $K^l(t):=\{\vh(t)\}$.
Finally, we compute $\vm^l=\vf^{-1}(\vh_r^l)=M_{an}(h_r^l)\frac{\vh_r^l}{h_r^l}$ and $\vm=\sum_{l=1}^n\omega^l\vm^l$.
 As an example, we simulated several hysteresis loops (Fig. \ref{Fig4}) for a material characterized by the anhysteretic function (\ref{Man}) with $m_s=1.23\cdot 10^6$ A/m, $A=50$ A/m, and represented by $N=20$ pseudoparticle types with $k^l=140(l-1)/(N-1)$ A/m, each having the same volume fraction $\omega^l=1/N$.

As was noted in \cite{B97}, it may be better to assume the magnetization of a pseudoparticle does not evolve independently but is influenced by the other particles. Hence, as the second essential modification of his model,  Bergqvist replaced the ``driving force" of this evolution, $\vh(t)$, by the ``effective" field $\vh(t)+\alpha \vm(t)$, where $\alpha$ is a material-dependent parameter. Such effective fields are often employed also in other models of hysteresis (see, e.g., \cite{Mgoyz,JA,LPA});  in \cite{Torre}  Della Torre presented an explanation of the interaction term $\alpha \vm(t)$ (see Ch. 4).
With this modification the convex time-dependent sets $K^l(t)$ should be replaced by
$$\begin{array}{c}K^l(t,\vm):=\{\vu \in {\mathbb R}^3\ :\\ |(k^l)^{-1}(\vu-\vh(t)-\alpha\vm)|\leq 1\}\end{array}$$
if $k^l\neq 0$ and $K^l(t,\vm):=\left\{\vh(t)+\alpha\vm\right\}$ otherwise. The internal variables $\vh_r^l$ are now solutions of the optimization problems
\beq \min_{\vu\in K^l(t,\vm)}\{S(\vu)-\check{\vm}^l\cdot\vu\},\label{optjm}\eeq
in which the constraints depend on the unknown solution itself, since $$\vm=\sum_{l=1}^N
\omega^l\vm^l=\sum_{l=1}^N
\omega^lM_{an}(h_r^l)\frac{\vh_r^l}{h_r^l}.$$
The implicit constraints in (\ref{optjm}) complicate the determination of the magnetizations  $\vm^l$ (such problems are equivalent to quasivariational inequalities). Nevertheless, an efficient iterative method can be proposed (see Section V).

Further modification of the model is needed to account for the known phenomenon of zero hysteresis loss in a rotational field at saturation \cite{Torre,Appino}. To describe this ``saturation property", it was suggested \cite{B97,B14,LZC15} to replace the constant intrinsic coercivity of each pseudoparticle, $k^l$, by a decreasing function $k^l(h_r^l)$ attaining zero at a saturation value $h_r^l=h_s$. In \cite{B14,LZC15}, however, only a ``monoparticle" model with the approximate update rule (\ref{Appr1}) has been considered. We suppose that in a composite model the lossless coherent rotation of all magnetization vectors $\vm^l$ is, probably, better described if, for all $l$, $k^l=k^l(m)$ which simultaneously drop to zero as the total magnetization $m$ reaches saturation. In this work, however, we consider for simplicity only constant $k^l$ values.

\section{Identification of the parameters in the model \label{Sec4}}
The practical implementation of the phenomenological model described above needs identification of the anhysteretic curve  $M_{an}$,  the material parameter $\alpha$, the coefficients $k^l$ (scalar or matrices, constant or dependent on $m$) and the corresponding weights $\omega^l$, $l=1,...,N$. Of course, the identification of parameters is needed also for other models of hysteresis and a variety of approaches have been proposed; the identification procedure depends on the model and the experimental data available.

\begin{figure}[!t]
\centering
\includegraphics[width=\linewidth,height=7.3cm]{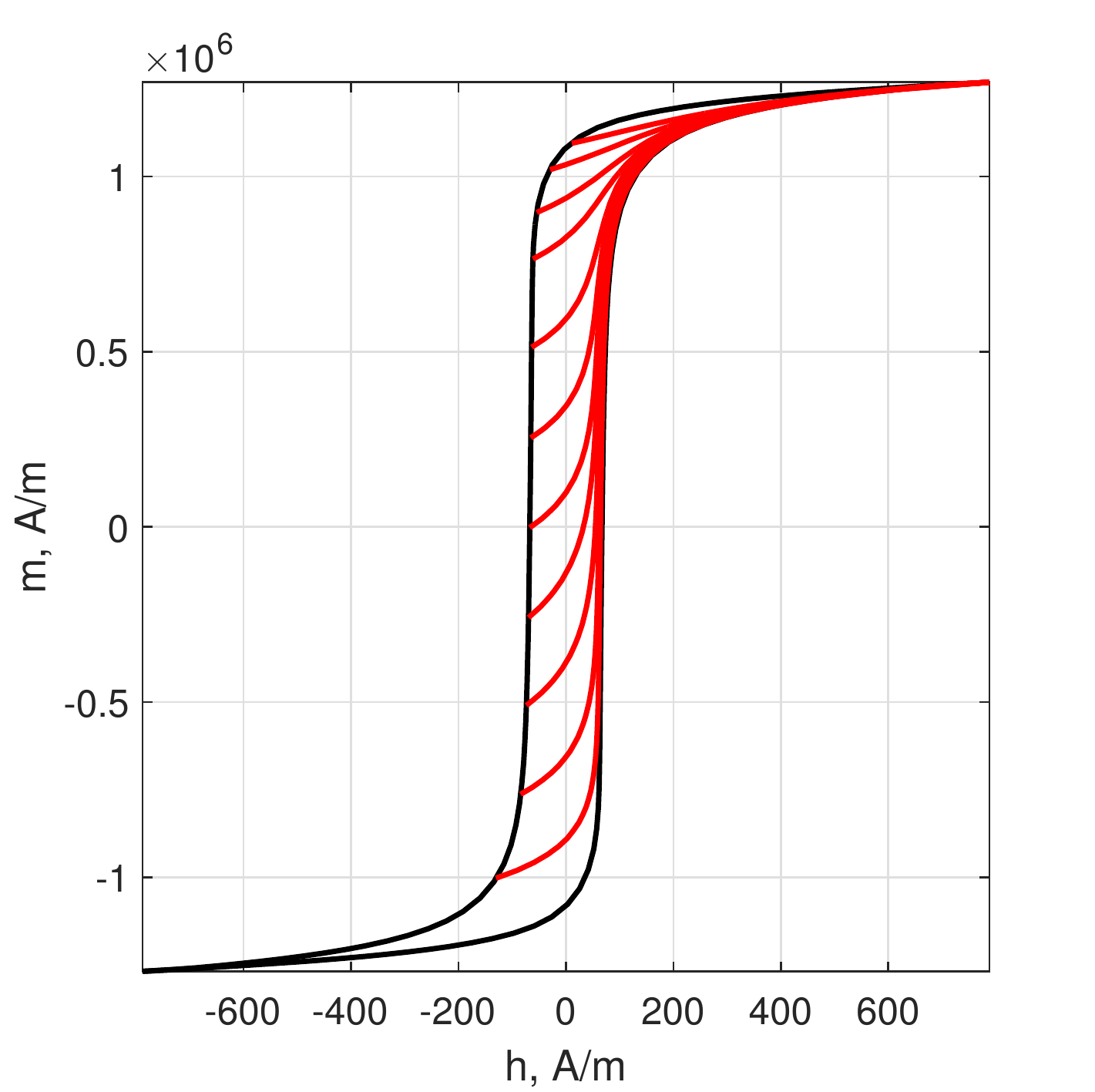}
\caption{Hysteresis curves, experimental.}\label{FigData}
\end{figure}

Here we present a consistent algorithm for the identification of the parameters in the model for a  nonoriented steel using the experimental major hysteresis loop and eleven first order reversal curves (FORCs) (the 1.8\% Si steel N3 in \cite{ZVis}, Fig. \ref{FigData}; see also \cite{ZInv}). Because of the symmetry, only the ascending part of the major loop was used; this curve was regarded as an additional FORC that starts at a strong negative magnetic field -800 A/m. We assume that the material is isotropic (in the anisotropic case similar measurements along the main material axes should be used for the identification) and, for simplicity, seek constant scalar values for $k^l$, $l=1,...,N$. Postulating that $N=41$ possible $k^l$ values are uniformly distributed from zero to 800 A/m with the step 20 A/m, we need to find only the appropriate weights $\omega^l$ to specify the distribution of the pseudoparticles.

No specific formulae was assumed for the anhysteretic function $M_{an}(|\vu|)$: this function was approximated by a cubic spline with the ``not-a-knot'' end conditions as follows. First, we determined the interval in which this function needs to be defined. Although the measured magnetic field did not exceed 800 A/m, the reversible fields $\vh_r^l$ driven by the effective field $\vh_{eff}=\vh+\alpha \vm$ can be out of this range. After several tests needed to estimate, at least crudely at first, the value of $\alpha$ (see below), we chose to approximate this function in the interval [0, 1750] A/m.  The anhysteretic function was then sought as a cubic spline with a fixed set of equidistant knots: $250i$ A/m, $i=0,1,...,7$. Since $M_{an}(0)=0$ is fixed, the spline is determined by its \textit{a priori} unknown values $M_i$ at the seven other knots. Note that in this experiment the fields can be regarded as scalar. In such a case let $h$ denote the field itself (which may be negative) and not its magnitude $|h|$ as above.

Let the anhysteretic function $M_{an}$, the particle fractions $\omega^l$, and the parameter $\alpha$ be given. At any point $\{h_0,m_0\}$ on the descending branch of the major hysteresis loop, we know the internal state $h_r^l$ of all pseudoparticle types: $h_r^l(h_0)=h_0+\alpha m_0+k^l$. If at this point the magnetic field starts to grow, the particle state evolution along the FORC $\{h,m(h)\}$ thus created can be described as $h_r^l(h)=\max\{h_r^l(h_0),h+\alpha m(h) -k^l\}$. The model prediction for the total magnetization along this curve, ${\cal M}(h)=\sum_{l=1}^N \omega^lM_{an}(h_r^l(h))$, can be compared to the measured $m^*(h)$ values for the same FORC. Our identification procedure finds the parameters $M_i$ of $M_{an}$, the weights $\omega^l$ and $\alpha$ minimizing the least-squares method residual, the sum $L$ of $[{\cal M}(h)-m^*(h)]^2$ over all measured points on all experimental FORCs. Efficient standard functions of Matlab were used in our three-level optimization algorithm as follows.
\begin{figure}[!h]
\centering
\includegraphics[width=\linewidth,height=7.3cm]{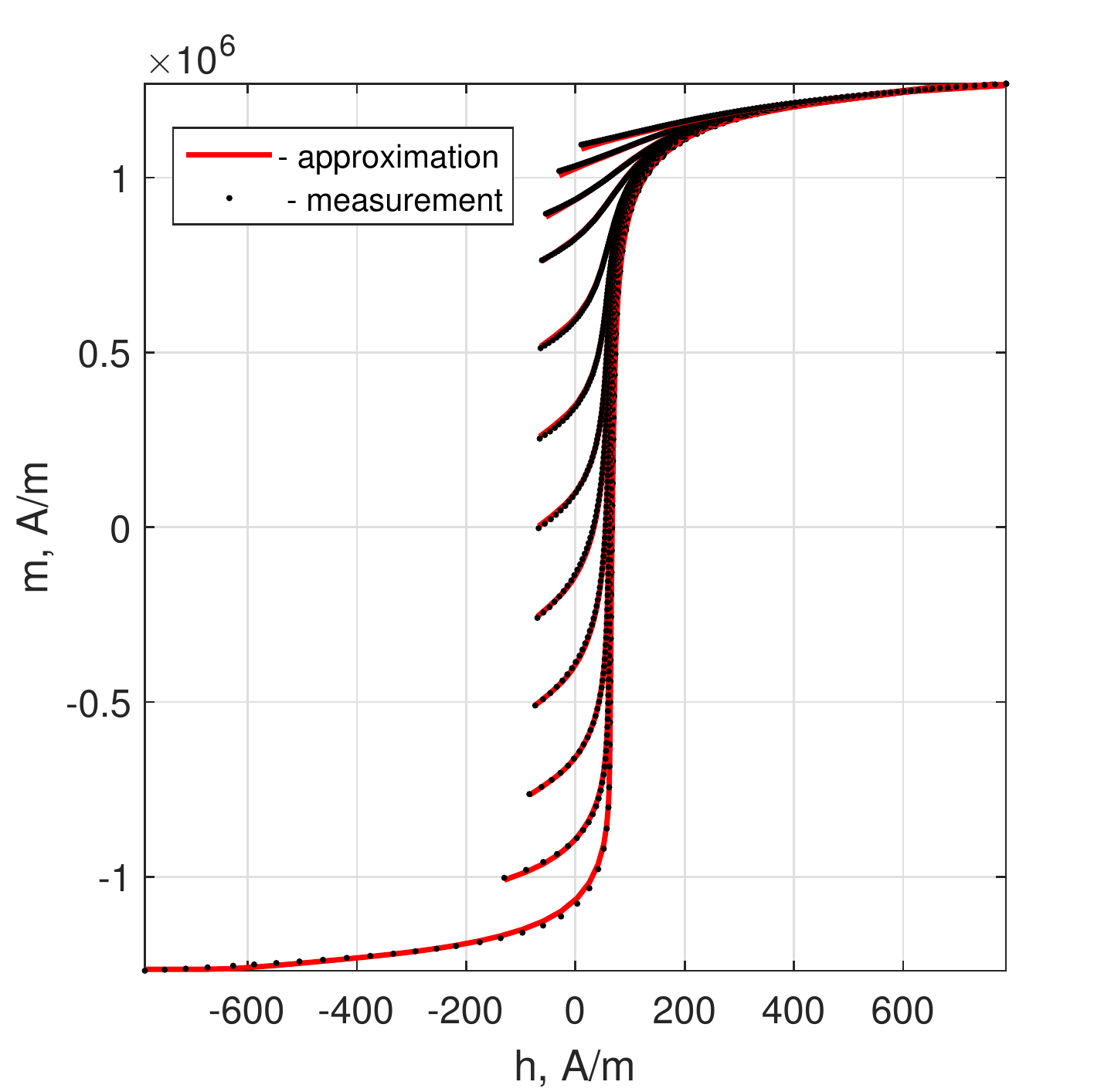}
 \caption{The least-squares fit of the FORCs.}\label{FigAppr}
\end{figure}
\begin{figure}[!t]\centering
\includegraphics[width=7.5cm,height=6cm]{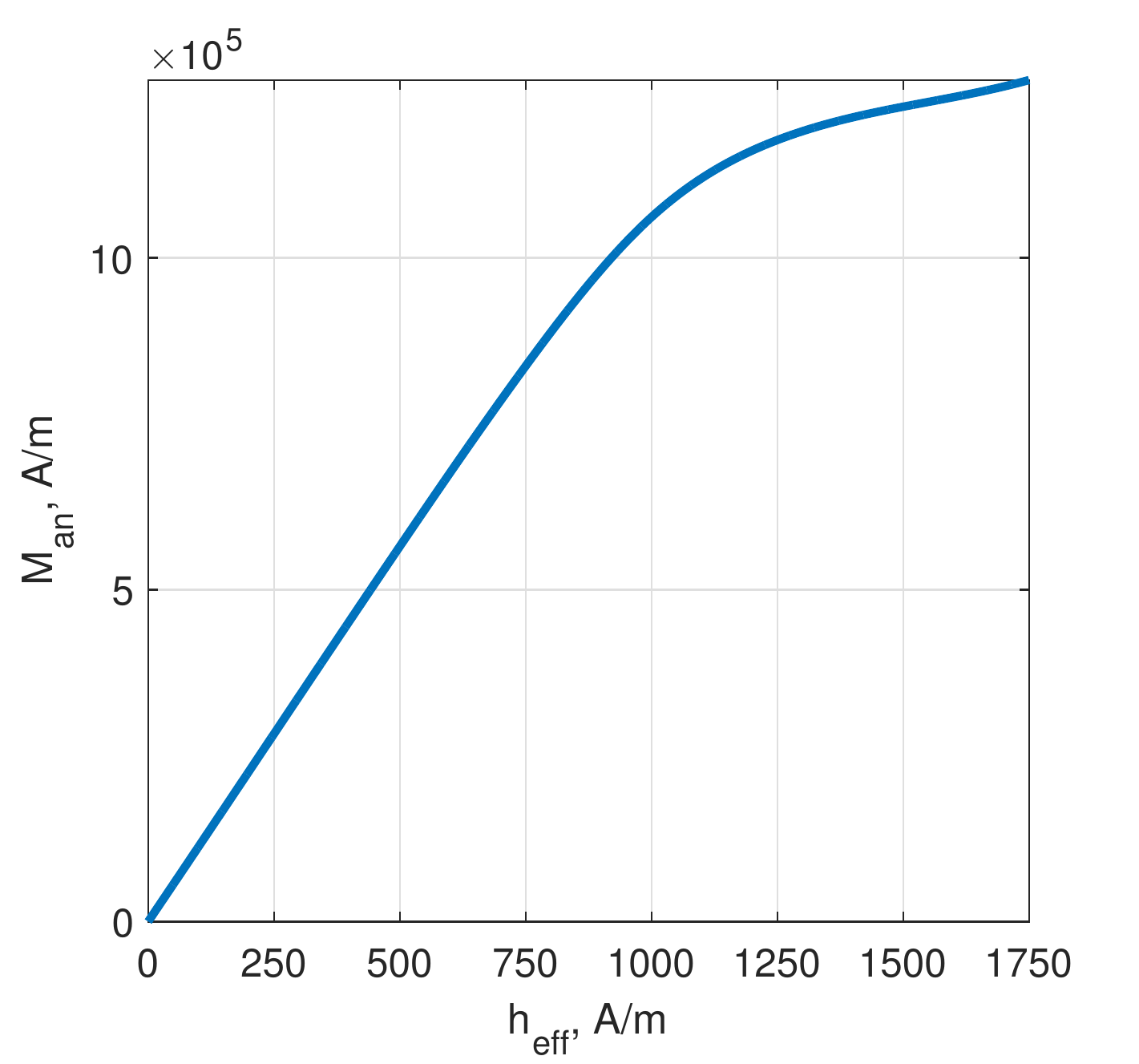}
\caption{Anhysteretic function $M_{an}$.}\label{FigMan}
\end{figure}
\begin{figure}[!h]\centering
  \includegraphics[width=7.5cm,height=6cm]{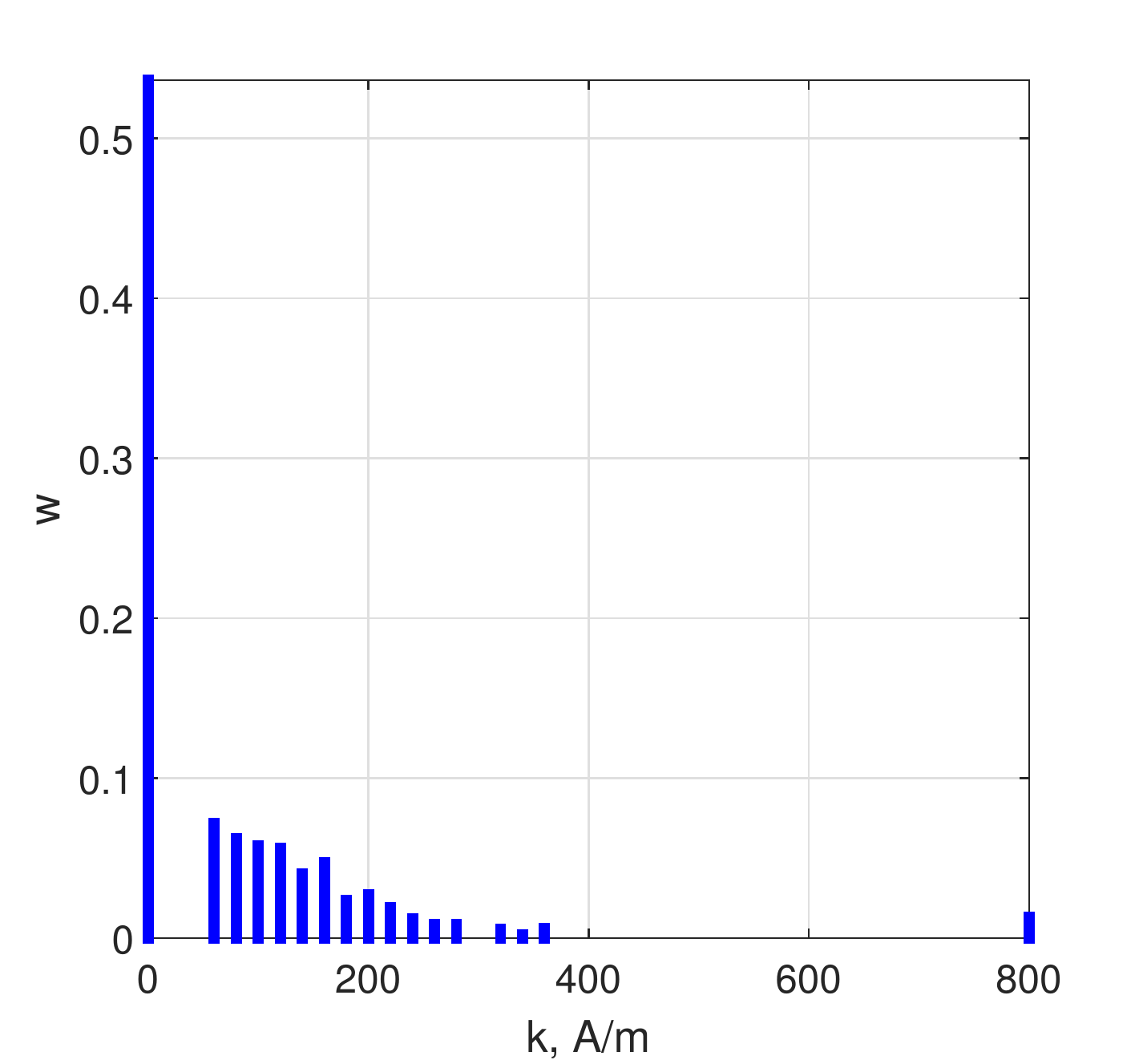}
  \caption{Weights $\omega(k)$; identification result.}\label{FigW}
\end{figure}

On the lowest level, the program determines the weights $\omega^l$ satisfying $\omega^l\geq 0,\ \sum_{l=1}^N\omega^l=1$ and minimizing the least-squares residual $L$  for given values of $M_i$ and $\alpha$. This is a quadratic programming problem solved by the \verb"lsqlin" function from the Matlab Optimization toolbox 7.2. On the next level, the Matlab function \verb"fminunc" realizes the unconstrained minimization over the set of $M_i$ for a given value of $\alpha$. Plotting the resulting function $L=L(\alpha)$ helped us to crudely estimate the position of the minimum and, on the upper level, this information was used by the Matlab function \verb"fminbnd" to find the optimal value of $\alpha$; the corresponding $M_i,\ \omega^l$ were thereby also obtained. Minimization of this function on a standard PC takes about 10 minutes.

Although we did not take into account any possible dependence of $k^l$ on $\vm$, a very good fit of the experimental curves has been obtained for $\alpha=8.8\cdot 10^{-4}$ (Fig. \ref{FigAppr}); the anhysteretic function $M_{an}$ and the weights $\omega^l$ are shown in Figs. \ref{FigMan} and \ref{FigW}, respectively.
We note that for only 17 of the 41 predefined values of $k^l$ did the identified volume fractions $\omega^l$ exceed $10^{-3}$. Only these pseudoparticles are taken into account in our finite element computations below; their total fraction is 0.9991. The results obtained show that for this material the reversible magnetization dominates (see Fig. \ref{FigW}) and there is a very significant reciprocal influence of pseudoparticles  via the $\alpha \vm$ term in $\vh_{eff}$ (see Figs. \ref{FigData} and \ref{FigMan}): without this term the $M_{an}(h)$ curve would be a much steeper line between the ascending and descending branches of the main hysteresis loop.

\section{A numerical scheme for a 2d hysteresis and eddy current problem}
Let us consider a long ferromagnetic cylinder, parallel to the $z$-axis and having a cross-section $\Omega$, carrying a transport current $I(t)$ and placed into a perpendicular uniform external field $\vh_e(t)$. The electric field $\ve(x,y,t)$ and the current density $\vj(x,y,t)$ are parallel to the $z$-axis; we can also choose the vector magnetic potential $\va(x,y,t)$ parallel to the $z$-axis (so that $\nabla\cdot\va=0$).   Hence,  these variables can be regarded as scalar and we will use the scalar notation $e,\ j$ and $a$ (which should not be confused with the absolute values of these vectors).
The vector fields $\vh(x,y,t),\vb(x,y,t)$ and $\vm(x,y,t)$ are parallel to the $xy$ plane.

We will use a 2d eddy current and magnetization problem formulation, similar to that proposed for 3d problems with hysteresis in \cite{dARTV}, but employ the Bergqvist model for magnetization discussed above. This model should be incorporated as a local constitutive relation
\beq \vm=\bm{M}[\vh_{eff}],\label{constM}\eeq
where $\vh_{eff}=\vh+\alpha\vm$ and $\vh=\frac{1}{\mu_0}\vb-\vm=\frac{1}{\mu_0}\vcurl a-\vm$. Here (and below) square brackets are used for the operator argument and $\vcurl a=(\partial_ya,-\partial_xa)$. The value $\bm{M}[\vh_{eff}]$ at a point depends on the history of $\vh_{eff}$ at this point: implicitly, the local operator $\bm{M}$ keeps track of the state of the internal variables $\vh^l_r$ and $\vm^l$.
 The density of the power loss due to magnetization is $W_H=\mu_0\sum_{l=1}^N\omega^l|k^l\dot{\vm}^l|$. Ohm's law $e= \rho j$, where $\rho$ is the resistivity, is another constitutive relation characterizing the material. The density of the eddy and transport current loss is $W_E=\rho j^2$.
We note that the so-called anomalous or excess loss, attributed to microscopic eddy currents caused by small, almost instantaneous jumps of domain walls (the Barkhausen noise), is not described by this model. As is shown in \cite{Zloss}, for the nonoriented steel considered in Section \ref{Sec4} this loss is small, at least for the magnetization in the rolling direction; for other magnetization directions the excess loss can, possibly, reach 10-20\% of the total loss, see \cite{Zloss1}.

For the specified geometry and gauge the electric field can be written as
$e=-\partial_t a+c(t),$ where the time-dependent constant $c(t)$ results from the parallel to the $z$-axis gradient of the scalar potential, $\nabla\Phi$ (see, e.g., \cite{P97}). This yields that
\beq \rho j=-\partial_t a+c(t),\eeq
where the unknown constant $c(t)$ is determined implicitly by the given transport current,
\beq\int_{\Omega}j\,d\vr =I(t).\eeq
In our numerical simulations we will assume that $I(t)=0$.

The vector potential can be represented as a sum, $a=a_e+a_j+a_m$, of the potentials associated with the external field, current density, and magnetization, respectively. Here $a_e=\mu_0(yh_{e,x}-xh_{e,y})$ and, see \cite{Jackson},
\begin{eqnarray} a_j[j]=\mu_0\int_{\Omega}G(\vr-\vr')j(\vr',t)\,d\vr',\label{aj}\\
a_m[\vm]=\mu_0\int_{\Omega}\nabla G(\vr-\vr')\times\vm(\vr',t)\,d\vr',\label{am}\end{eqnarray}
where $\nabla$ is the gradient with respect to $\vr=(x,y)$
and $G(\vr)=\frac{1}{2\pi}\ln\frac{1}{|\vr|}$ is the Green's function.

Correspondingly, the magnetic field can be represented as the sum:
$\vh=\vh_e+\vh_j+\vh_m,$ where  \begin{align*} \vh_j[j]=&\int_{\Omega}\nabla G(\vr-\vr')\times j(\vr',t)\,d\vr'=\\&\left( \int_{\Omega}j(\vr',t)\partial_{y}G(\vr-\vr')\,d\vr',\right.\\& \left. -\int_{\Omega}j(\vr',t)\partial_{x}G(\vr-\vr')\,d\vr' \right)\end{align*}
and, see \cite{bofm},
\begin{align} \vh_m[\vm]=&\frac{1}{\mu_0}\vb_m-\vm=\nonumber\\ &\nabla\int_{\Omega}\nabla\cdot\left[G(\vr-\vr')\vm(\vr',t)\right]\,d\vr'.
\label{bm}\end{align}
The main unknowns in this model are $\vm(\vr,t)$ and $j(\vr,t)$: provided these variables are found, $\vh$ and $\vb$ can be also calculated; $c(t)$ is an auxiliary unknown. After discretization in time, the problem to be solved on each time level $n$ consists of three linear
equations,
\begin{align}\tau\rho j^{n}&+a_j[j^{n}]+a_m[\vm^{n}]-\tau{c}^{n}=\nonumber\\&a_e^{n-1}-a_e^{n}+a_j[j^{n-1}]+
a_m[\vm^{n-1}],\label{it1}\\
&\int_{\Omega}j^{n}\,d\vr=I^n,\label{it2}\\
&\vh^{n}_{eff}=\vh_e^n+\vh_j[j^{n}]+\vh_m[\vm^{n}]+\alpha\vm^{n},\label{ith}
\end{align}
where $\tau$ is the time step, supplemented by the nonlinear relationship,
\beq\vm^{n}=\bm{M}[\vh^{n}_{eff}]\label{m_up}.\eeq
Our iterative scheme was based on the representation
\beq \vm^{n,k+1}=\bm{M}[\vh_{eff}^{n,k}]+D[\vh_{eff}^{n,k}]\left(\vh_{eff}^{n,k+1}-\vh_{eff}^{n,k}\right),\label{it3}\eeq
where $k$ is the iteration number and
$D[\vu]$ is, at each point of $\Omega$, the $2\times 2$ matrix of partial derivatives,
$$D[\vu]=\left(\begin{array}{cc}
\partial_{u_x} M_x[\vu]&\partial_{u_y} M_x[\vu]\\
\partial_{u_x} M_y[\vu]&\partial_{u_y} M_y[\vu]
\end{array}\right).$$
Since the analytical calculation of these derivatives is difficult, we used their numerical approximations:
replacing $u_x$ by $u_x\pm \Delta$ and keeping $u_y$ unchanged, we used the central difference to estimate $\partial_{u_x} \bm{M}[\vu]$; similarly for $\partial_{u_y} \bm{M}[\vu]$.

Substituting (\ref{it3})  into (\ref{it1})--(\ref{ith}) yields a linear system for $\vh_{eff}^{n,k+1},\ j^{n,k+1},\ {c}^{n,k+1}$. These iterations should be repeated until convergence of $\vh_{eff}^{n,k}$ with a given tolerance; the magnetization and magnetic field in the magnetic material are then found as
$\vm^n=\bm{M}[\vh_{eff}^n]$ and
$\vh^{n}=\vh^n_{eff}-\alpha\vm^n$, respectively.

The described iterative procedure has been applied to the finite element approximation of this problem.
We triangulated the domain $\Omega$ and used piecewise constant approximations for all variables, $j,\ a,\ \vh$ and $\vm$.
The finite element
approximation of the integral operator equation (\ref{it1}) involves the computation of matrices with the entries, see (\ref{aj}), (\ref{am}),
\begin{eqnarray} L_{e,e'}=\int_e\int_{e'}G(\vr-\vr')d\vr'\,d\vr,\label{M_int}\\ L^x_{e,e'}=\int_e\int_{e'}
\partial_xG(\vr-\vr')d\vr'\,d\vr,\\ L^y_{e,e'}=\int_e\int_{e'}\partial_yG(\vr-\vr')d\vr'\,d\vr.\label{Mx_int}\end{eqnarray}
for each pair of triangles $e,\,e'$.
Only the matrices $L^x,\ L^y$ defined above are needed to find $\vh_j[j]$ in (\ref{ith}). However, to compute
the $\vh_m[\vm]$ part of $\vh_{eff}$ one needs, see (\ref{bm}), matrices with entries
\begin{eqnarray} L^{xx}_{e,e'}=\int_e\int_{e'}\partial^2_{xx}G(\vr-\vr')d\vr'\,d\vr,\label{Mxx_int}\\
L^{yy}_{e,e'}=\int_e\int_{e'}\partial^2_{yy}G(\vr-\vr')d\vr'\,d\vr,\\
L^{xy}_{e,e'}=L^{yx}_{e,e'}=\int_e\int_{e'}\partial^2_{xy}G(\vr-\vr')d\vr'\,d\vr.
\label{Myy_int}\end{eqnarray}
Some of the integrals in (\ref{M_int})--(\ref{Myy_int}) are singular; their computation is described in Appendix B. We note that the matrices $L,\ L^{xx},\ L^{xy}$ and $L^{yy}$
are symmetric, whereas the $L^x,\ L^y$ are antisymmetric.

To approximate the matrix $D$ in each element we used $\Delta =2\cdot 10^{-6}\max(1,h)$. Convergence of the iterations was stable and fast: three-five iterations per time level ensured convergence with the relative tolerance $10^{-6}$ (in the $L^1$ norm) in all our simulations.

\section{Simulation results}
In all numerical simulations below we assumed the material is isotropic and used the hysteresis model parameters identified for a nonoriented steel in Section \ref{Sec4}.
As our first example, we considered a cylinder with a circular cross-section $\Omega=\{\vr\ :\ r<r_0\}$. We made the eddy current negligible by choosing a very high resistivity $\rho$. Then, for a material with constant magnetic permeability $\mu$, the magnetic field can be expressed via the scalar magnetic potential, $\vh=-\nabla \psi$,  satisfying the Laplace equation
$\frac{1}{r}\frac{\partial}{\partial r}\left(r\frac{\partial\psi}{\partial r}\right)+\frac{1}{r^2}\frac{\partial^2\psi}{\partial \theta^2}=0$
inside and outside the domain $\Omega$ with the interface conditions $$\psi|_{r_0-}=\psi|_{r_0+},\ \ \ \left.\mu\frac{\partial\psi}{\partial r}\right|_{r_0-}=\left.\mu_0\frac{\partial\psi}{\partial r}\right|_{r_0+}$$ and the boundary condition $-\nabla \psi\rightarrow\vh_e$ as $r\rightarrow \infty$. Solving this problem by separation of variables, one finds that inside the cylinder the fields $\vh$ and $\vm$ are uniform:
\beq\vh=\frac{2\vh_e}{\mu_r +1},\ \ \vm=\frac{2\vh_e(\mu_r-1)}{\mu_r +1},\label{fields}\eeq
where $\mu_r=\mu/\mu_0$.
For a circular ferromagnetic cylinder we now assume the virgin initial state ($\vm^l|_{t=0}=0$ for all $l$) and a uniform external field $\vh_e(t)$ growing from zero monotonically in a fixed direction. In this case $\vh$ and $\vm$ are also uniform in $\Omega$ but obey (\ref{fields}) with an unknown relative permeability $\mu_r$ varying with $h_e$. For this unidirectional situation the model (\ref{constM}) employed predicts
\beq m=\sum_{l=1}^N\omega^lM_{an}([h+\alpha m-k^l]_+),\label{nl}\eeq
where $u_+=\max\{u,0\}$. Substituting relations (\ref{fields}) into (\ref{nl}) we arrive at a nonlinear algebraic equation for $\mu_r$,  which is easy to solve numerically; this determines $h$ and $m$ for  any given $h_e$.

This solution does not depend on $r_0$ and was used as a partial test for our  finite element simulations.
We set $\vh_{e}=(10^3t,\,0)$ A/m and used two finite element meshes (742 and 2436 triangles) to compute the solution at $t=100$ s. For the crude mesh the relative errors of our finite element solution (in the $L^1$ norm) did not exceed $\delta_h=0.08\%$ and $\delta_m=0.21\%$ for $\vh$ and $\vm$, respectively; for the finer mesh we obtained $\delta_h=0.05\%$ and $\delta_m=0.18\%$. The moderate accuracy gain for $\vm$ for the finer mesh probably indicates that a non-negligible part of the error is induced by the use of numerical integration in (\ref{M_int})--(\ref{Myy_int}). We found that in this example the accuracy was practically independent of the constant time step $\tau$; such a phenomenon is often observed in modeling ``simple regimes" of rate-independent models (see, e.g., \cite{BP14}, p. 1010) and such is the employed model without the eddy current.
The average number of iterations per time level did not depend on the mesh and increased from 3.7 for $\tau=5$ s to 4.5 for $\tau=50$ s.

\begin{figure}[!h]
\centering
\includegraphics[width=8.5cm]{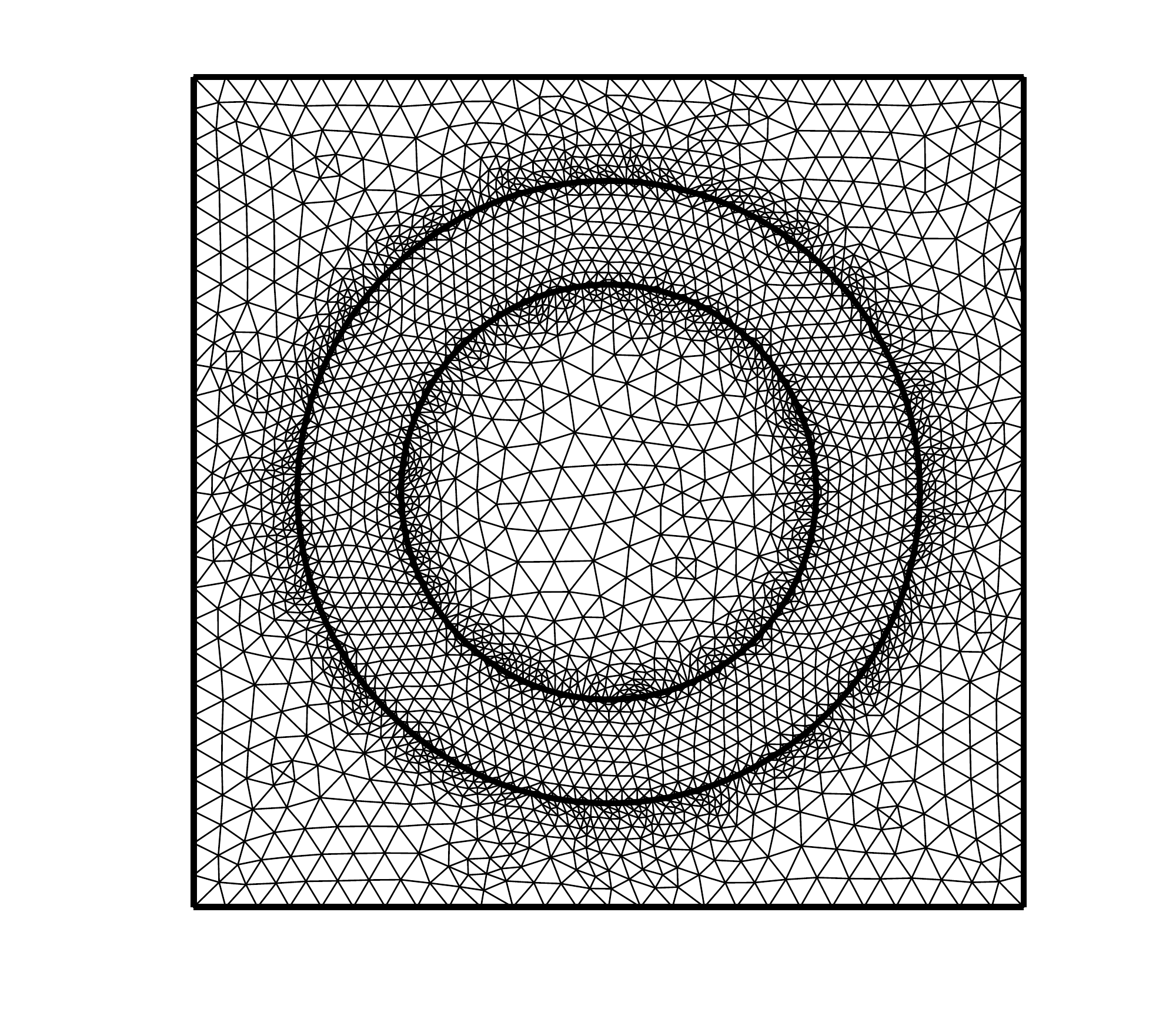}
\caption{Finite element mesh.}\label{Fig_mesh}
\end{figure}

\begin{figure*}[!t]
\centering
\includegraphics[width=7.6cm]{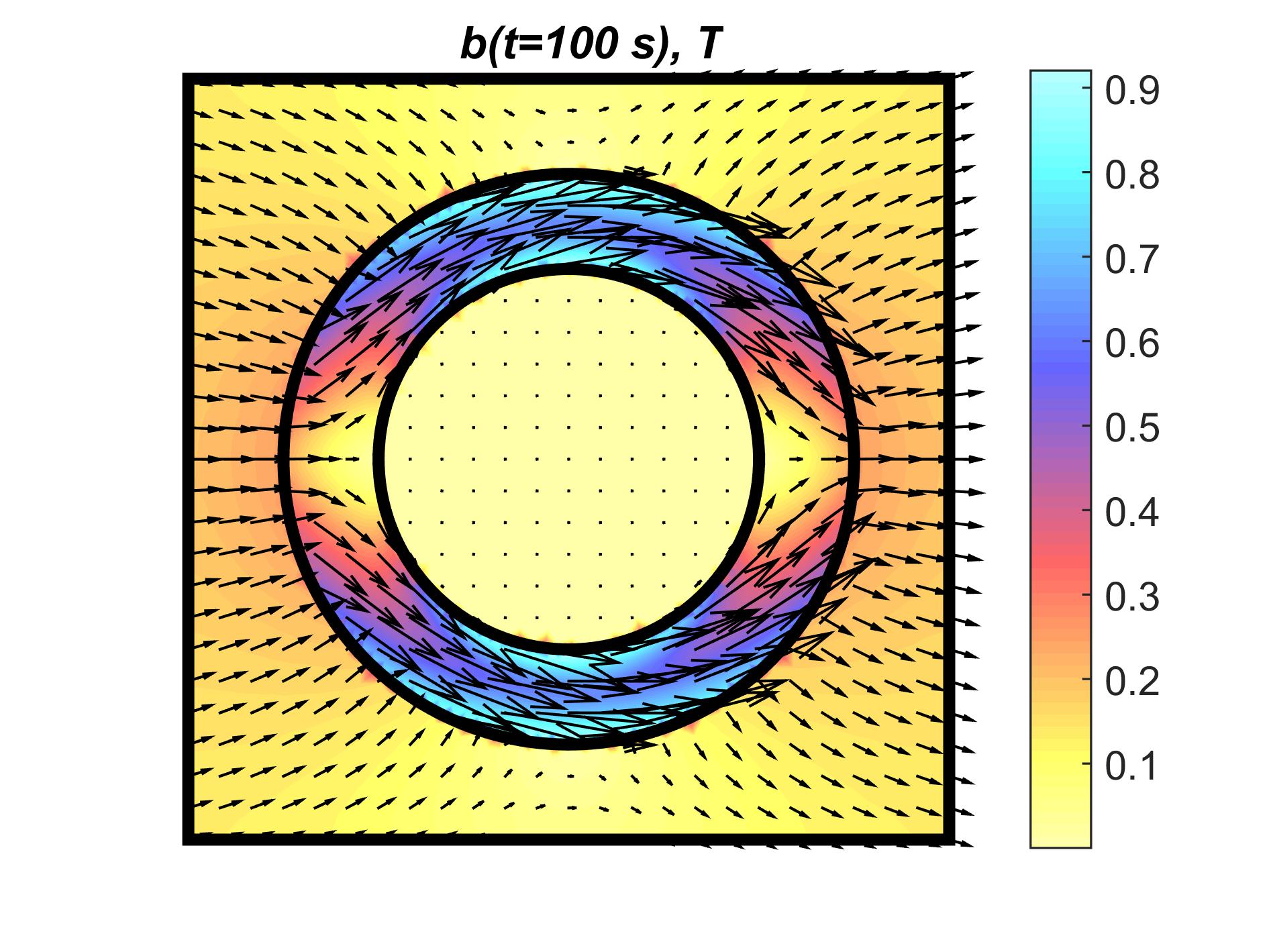}\includegraphics[width=7.6cm]{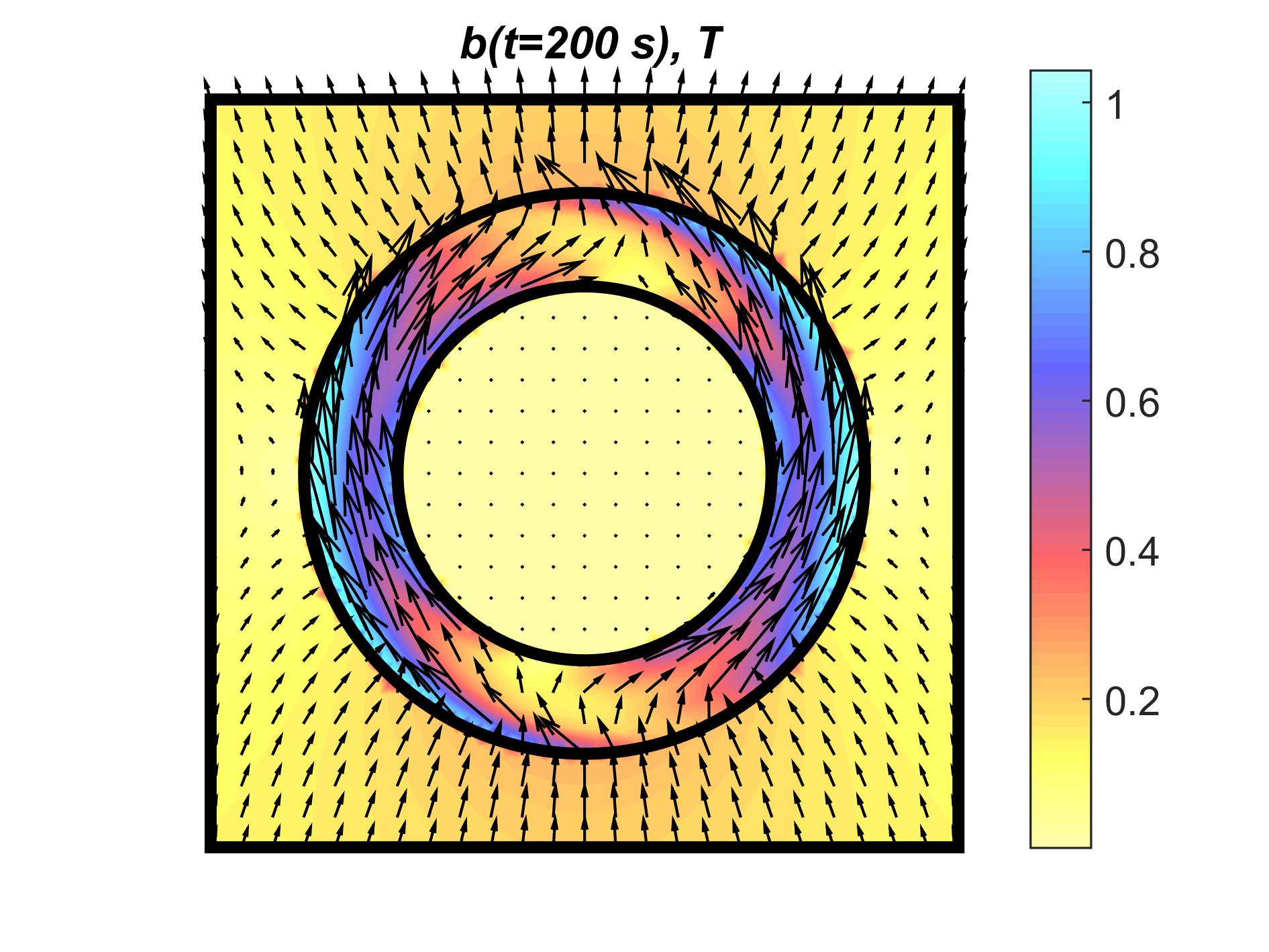}\\
\includegraphics[width=16.8cm]{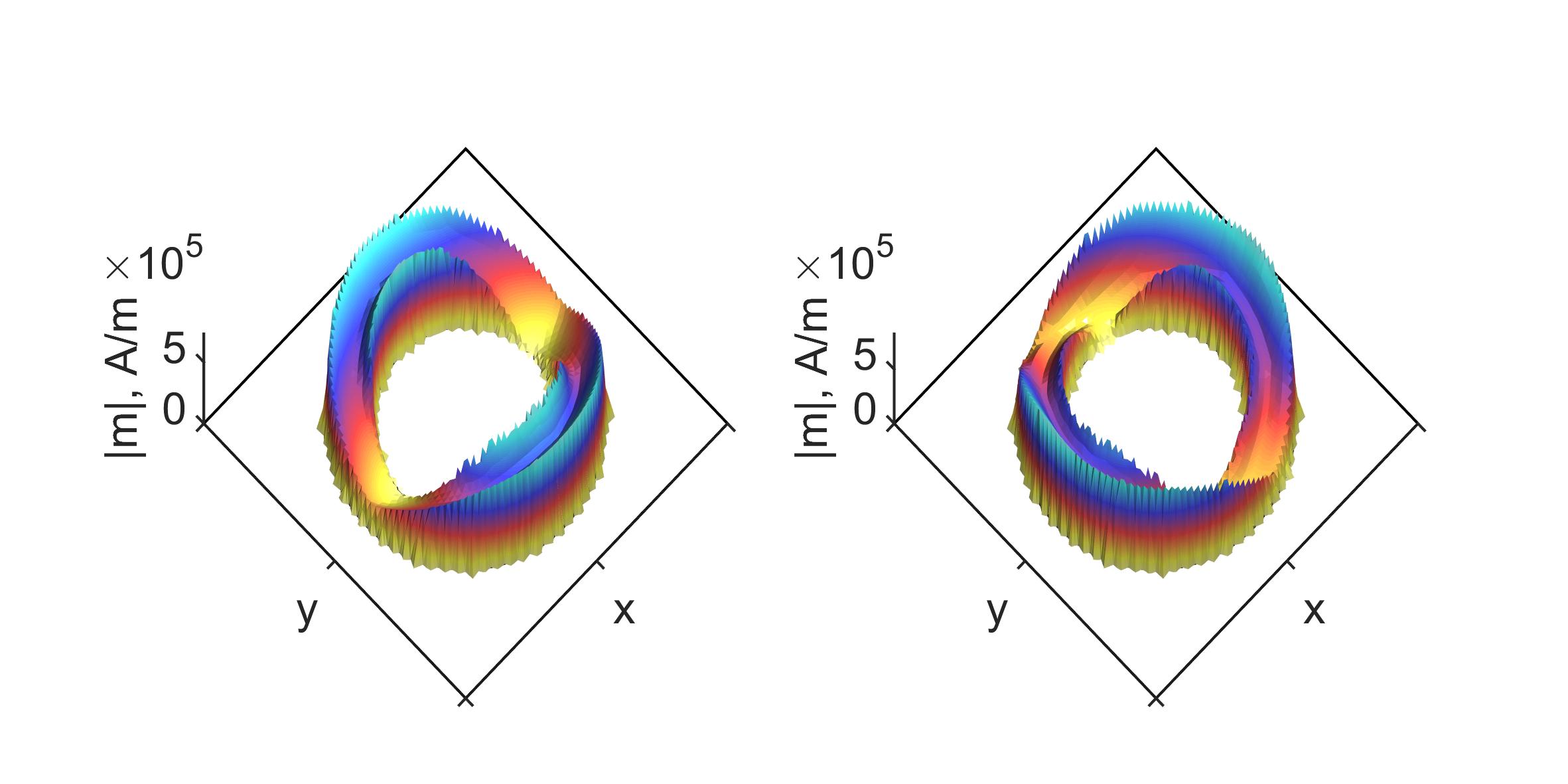}\\
\includegraphics[width=16.8cm]{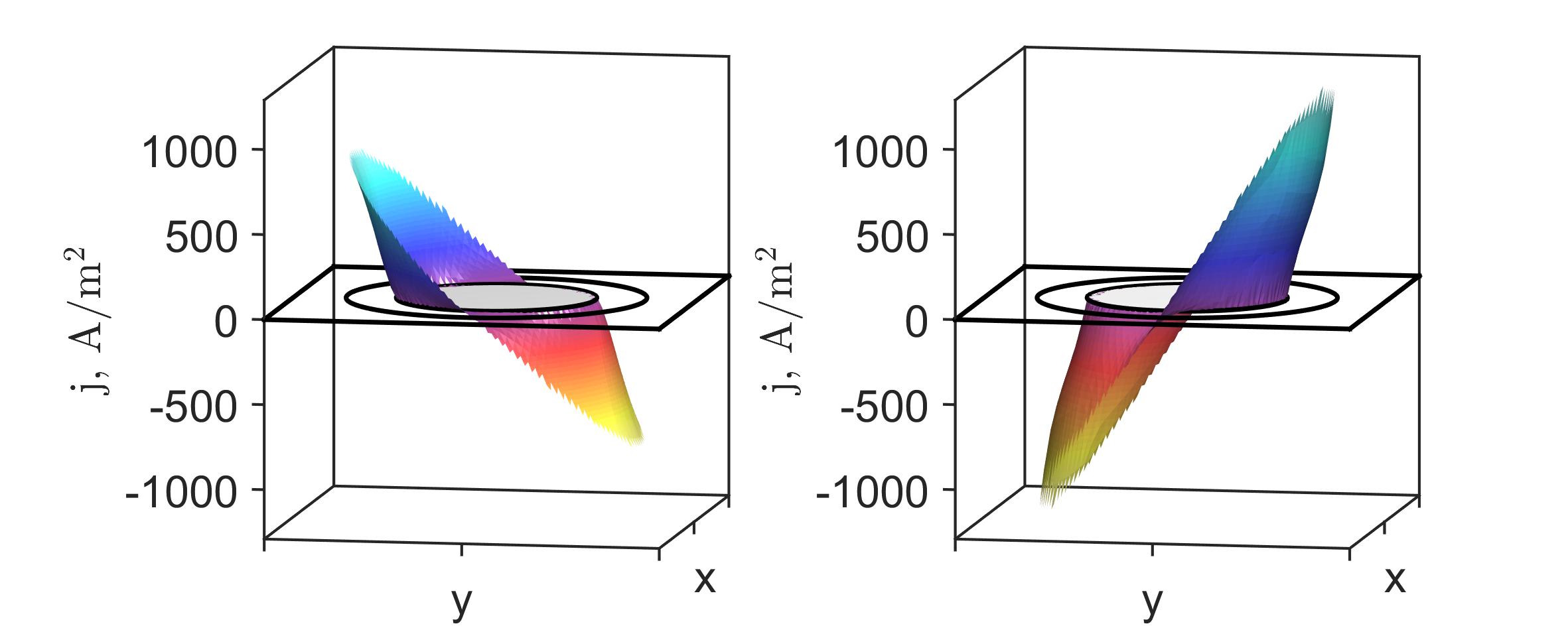}
\caption{Numerical simulation. Left: solution for $t=100$ s, right: for $t=200$ s. Top: magnetic induction; middle: magnetization $|\vm|$; bottom: eddy current density. }\label{Fig_ring}
\end{figure*}

Our second example is a hollow ferromagnetic cylinder with the cross-section $r_1\leq r\leq r_2$; such a configuration can be employed for magnetic shielding. If the eddy current can be neglected and the magnetic permeability of the material is constant, the problem can be solved analytically \cite{Bess}. In this case the magnetic field inside the hole is uniform:
$$\vh=\frac{4\mu_r\vh_e}{(\mu_r+1)^2-(r_1/r_2)^2(\mu_r-1)^2}.$$
 Assuming the steel resistivity $\rho=0.43\, \mu\Omega\cdot$m (see \cite{ZVis}) and taking $r_1=0.1$ m, $r_2=0.15$ m we solved the magnetization problem taking both the eddy current and hysteresis into account.  We now triangulated a larger than the cross-section domain, the square $-0.2\leq x,y\leq 0.2$ m, see Fig. \ref{Fig_mesh}. The mesh contains 6424 triangles; 2352 of them belong to the ferromagnetic domain.
 Although our numerical algorithm needs only the latter elements, to find the magnetic induction also outside the ferromagnet we computed the elements of matrices (\ref{M_int})--(\ref{Myy_int}) for all triangle pairs $e,\,e'$, where at least one of the triangles belongs to the magnetic domain (the remaining elements of these matrices can be set to zero). We set $\vh_e=(10^3t,0)$ A/m for the first 100 s and assumed that in the next 100 s the external field rotates $90^{\circ}$ counter clockwise with a constant angular velocity, its magnitude remaining at $10^5$ A/m. The time steps $\tau=10$ s and $\tau=2.5$ s were used, respectively, for these two time intervals; on average, convergence with the relative tolerance $10^{-6}$ was achieved in five iterations per time level. Numerical simulation results for $t = 100$ s and  $t = 200$ s are shown in Fig. \ref{Fig_ring}.
\clearpage
\noindent A further decrease of the time steps does not change these results.

The eddy current density is not negligible now but,
according to our computation, for the assumed external field variation rate the power of the eddy current loss per unit of length,
$p_j=\int_{\Omega}\rho j^2 d\vr$, is smaller than the power of the magnetization
dissipation $p_m=\mu_0\int_{\Omega}\left\{\sum_l k^l|\dot{\vm}^l|\right\} d\vr$. Thus, in the end of the linear growth of $h_e$ ($t=100$ s) we obtained $p_j=2.4\cdot 10^{-3}$ W/m and $p_m=15\cdot 10^{-3}$ W/m; at $t=200$ s (rotating $\vh_e$) we obtained $p_j=5.0\cdot 10^{-3}$ W/m and $p_m=20\cdot 10^{-3}$ W/m. The corresponding losses during the whole time interval $0<t<200$ s are 0.70 J/m and 2.3 J/m, respectively. It may be noted that not only the eddy current density, but also the related part of the total loss should increase with the ramping rate of the external field. We found also that the magnetic field inside the cylindrical hole does not exceed 0.14\% of $h_e$ (the shielding effect). Further simulations showed that the pronounced non-monotonicity of the magnetization $\vm(\vr,t)$ in the radial directions (see Fig. \ref{Fig_ring}, middle) is caused by the eddy current. Suppressing this current by choosing a high resistivity value makes the solution monotonic radially; qualitatively, for $t\leq 100$ s the solution without the eddy current resembles and corresponds to a similar shielding as the analytical solution for an appropriate constant permeability, $\mu_r\approx 7\cdot 10^3$. However, since the magnetic field inside the magnet is not uniform, in the hysteretic (and even simply in the nonlinear) case obtaining an analytical solution (as we did in the previous example) is not possible. The numerical simulation showed that also the magnetic field inside the hole is not uniform. Finally, we note that, during the rotation of the external field, the magnetization of the inner ring layers lags behind (Fig. \ref{Fig_ring}, middle-right), which is the hysteresis effect.

It should be noted that the magnetization model employed here is complicated and cannot compete with much simpler models in terms of the computation time. Even after the matrices (\ref{M_int})--(\ref{Myy_int}) were computed, our Matlab program still needed several hours to solve examples like these on a standard PC. The main elements, determining the efficiency of our computations are: the integral vector potential formulation, the inner iterations for updating the magnetization of pseudoparticles in each finite element, and the
outer iterations for solving (\ref{it1})--(\ref{m_up}) using (\ref{it3}).

Integral formulations like this (we followed \cite{dARTV}) are often employed for solving electrical engineering problems in general, and eddy current problems in particular. Their main advantage is that all computations are confined to the area occupied by the conducting and/or magnetic materials; but a disadvantage is the linear systems with dense matrices. Overall, such formulations are competitive with other formulations used for finite element approximations.

The inner iterations converged in 2--3 iterations, much faster than in \cite{FLavHenr2013} and, as was shown above (see Fig. 3), the faster approach based on the explicit approximation (\ref{Appr1}) can be inexact.

As is well known, for nonlinear materials with high differential susceptibility
values it is difficult to obtain good convergence of iterations in finite element
simulations, especially, if the Newton method cannot be employed. This is the case for the model employed here: in the Newton-like method that we used the necessary derivatives could only be numerically approximated and, for steel in our simulations, the maximal susceptibility exceeded $10^5$. Nevertheless, we were able to reach convergence in five iterations per time level in the outer cycle of our scheme, which is a fast convergence.

\section{Conclusions}
Like most existing macroscopic models for ferromagnetic hysteresis, the quasi-static model proposed by Bergqvist in \cite{B97} is phenomenological. However, it is based upon consistent energy arguments and a clear albeit simplified physical picture of the dry-friction like pinning of the domain walls. This model is naturally vectorial, has a variational formulation convenient for numerical simulations, and can be incorporated  into a finite element code as a local constitutive relation with memory. As an example we considered a problem, where both the magnetization and eddy current were taken into account.

In this work we tried to clarify the mathematical derivation of the variational formulation, extended it to the anisotropic case, proposed an efficient numerical method based on this formulation, and also demonstrated that the usually employed approximation, which turns the model into a play hysteron model, can be inaccurate. We showed that this approximation is not a possible version of the dry friction law, as is typically assumed, but a replacement by an alternative assumption, not related to dry friction, and determining a different direction of the system's evolution in the vectorial case.

The model has sufficient degrees of freedom to be fitted to hysteretic behavior of different  materials; here  we presented a method for the identification of the parameters in this model using a set of experimental FORCs. Another advantage of this model is its ability to predict both the stored and dissipated energies at any moment in time. These properties make the model highly attractive; its further comparison to experiments would be desirable.


\appendices
\section{Newton's method for (\ref{2dopt})}
Let $\vu(\phi)=\vh(t)+k[\cos \phi,\sin \phi]^T$. Then $g(\phi)=S(\vu(\phi))-\vmc\cdot\vu(\phi)$ and we solve $g'(\phi)=0$ iteratively,  $\phi_{n+1}=\phi_n-g'(\phi_n)/g''(\phi_n)$, with $\phi_0$ being the direction of $k^{-1}(\check{\vh}_r-\vh(t))$. Here
$$g'=\{\vnabla S(\vu)-\vmc\}\cdot \vu'=\left\{\frac{M(u)}{u}\vu-\vmc\right\}\cdot \vu',$$
\begin{align*}g''=&\left\{\vnabla\left(\frac{M(u)}{u}\right)\cdot\vu'\right\}\vu\cdot\vu'+\frac{M(u)}{u}\vu'\cdot\vu'+\\
&\left\{\frac{M(u)}{u}\vu-\vmc\right\}\cdot \vu'',\end{align*}
where $\vu'=k[-\sin \phi,\cos \phi]^T,\ \vu''=-k[\cos \phi,\sin \phi]^T$ and $$\vnabla\left(\frac{M(u)}{u}\right)=\frac{d}{du}\left(\frac{M(u)}{u}\right)\frac{\vu}{u}=\frac{uM'-M}{u^2}\frac{\vu}{u}.$$ %

\section{Computation of the integrals (\ref{M_int})--(\ref{Myy_int})}
To compute the integrals (\ref{M_int})--(\ref{Myy_int}) for any distant pair of triangles $e,\, e'$ we used a symmetric seven point quadrature rule for triangles, exact for all polynomials of degree five \cite{Bur} (the triangles were regarded as distant if the distance between their centers exceeded several characteristic triangle sizes). For the coinciding, touching, and also close to each other triangles $e$ and $e'$ we employed the representation (\cite{Kat}, p. 305) $G(\vr)=\Delta U(\vr)-\frac{1}{2\pi},$ where $U(\vr)=-\frac{1}{8\pi}r^2\ln(r)$; this yields
\begin{align*}&\int_e\int_{e'}G(\vr-\vr')d\vr'\,d\vr=\\ \ &\int_e\int_{e'}\Delta U(\vr-\vr')d\vr'\,d\vr-\frac{|e||e'|}{2\pi}.\end{align*}
In addition, we have that
\begin{eqnarray*}\int_e\int_{e'}\Delta U(\vr-\vr')d\vr'\,d\vr=\\-\int_e\nabla\cdot\int_{e'}\nabla' U(\vr-\vr')d\vr'\,d\vr=\\
-\oint_{\partial e}\vn\cdot\left\{\int_{e'}\nabla'U(\vr-\vr')d\vr'\right\}dl=\\-\oint_{\partial e}\left\{\int_{e'}\nabla'\cdot[U(\vr-\vr')\vn]d\vr'\right\}dl
=\\-\oint_{\partial e}\oint_{\partial e'}U(\vr-\vr')\vn\cdot\vn'\, dl'dl,\end{eqnarray*}
where $\vn$ is the outward unit normal to the boundary $\partial e$ of $e$.
Similarly,  e.g., \begin{eqnarray*}\int_e\int_{e'}\partial^2_{xx}G(\vr-\vr')d\vr'\,d\vr=\\-\oint_{\partial e}\oint_{\partial e'}\partial^2_{xx}U(\vr-\vr')\vn\cdot\vn'\, dl'dl,\end{eqnarray*}
so all double surface integrals can be written as a sum of integrals over pairs of edges of the triangles.
We note that all such integrals of $U$ and its first derivatives
$$\partial_xU=-\frac{(\ln(r^2)+1)x}{8\pi},\ \ \partial_yU=-\frac{(\ln(r^2)+1)y}{8\pi}$$
are regular. These integrals were computed numerically with a controlled accuracy.  The integrals of
\begin{eqnarray*}\partial^2_{xy}U=-\frac{xy}{4\pi r^2},\\  \partial_{xx}^2U=-\frac{1}{8\pi}\left(\frac{2x^2}{r^2}+\ln(r^2)+1\right),\\ \partial^2_{yy}U=-\frac{1}{8\pi}\left(\frac{2y^2}{r^2}+\ln(r^2)+1\right)\end{eqnarray*}
were calculated analytically if the two edges coincided or had a common vertex (we used the Matlab Symbolic toolbox 6.1 in the latter case); otherwise these integrals are also regular and were computed numerically.

\begin{thebibliography}{11}
\bibitem{Mgoyz} I. D. Mayergoyz, \textit{Mathematical Models of Hysteresis} (Springer, NY, 1991).
\bibitem{JA} D. C. Jiles and D. Atherton, Theory of ferromagnetic hysteresis. \textit{J. Magn. Magn. Mat.} \textbf{61} (1986) 48-60.
\bibitem{Carp} K. H. Carpenter, A differential equation approach to minor loops in
the Jiles-Atherton hysteresis model. \textit{IEEE Trans. Magn.} \textbf{27} (1991) 4404-4406.
\bibitem{MZ} D. Miljavec and B. Zidari\v{c}, Introducing a domain flexing function in the Jiles-Atherton hysteresis model. \textit{J. Magn. Magn. Mat.} \textbf{320} (2008) 763-768.
\bibitem{Zirka} S. E. Zirka, Yu. I. Moroz, R. G. Harrison and K. Chwastek,  On physical aspects of the Jiles-Atherton hysteresis models. \textit{J. Appl. Phys.} \textbf{112} (2012) 043916.
\bibitem{FrMgoyz} G. Friedman and I. D. Mayergoyz, Hysteresis energy losses in media described by vector Preisach model. \textit{IEEE Trans. Magn.} \textbf{34} (1998) 1270-1272.
\bibitem{B97} A. Bergqvist, Magnetic vector hysteresis model with dry-friction like pinning. \textit{Physica} B \textbf{233} (1997) 342-347.
\bibitem{B97b} A. Bergqvist, A. Lundgren and G. Engdahl, Experimental testing of an anisotropic vector hysteresis model. \textit{IEEE Trans. Magnetics} \textbf{33} (1997) 4152-4154.
\bibitem{KrahB04} J. H. Krah and A. Bergqvist, Numerical optimization of a hysteresis model. \textit{Physica} B \textbf{343} (2004) 35-38.
\bibitem{HenrNicHam2006} F. Henrotte, A. Nicolet and K. Hameyer, An energy-based vector hysteresis model for ferromagnetic materials. \textit{COMPEL} \textbf{25} (2006) 71-80.
\bibitem{HenrHam2006} F. Henrotte and K. Hameyer, A dynamical vector hysteresis model based on an energy approach. \textit{IEEE Trans. Magnetics} \textbf{42} (2006) 899-902.
\bibitem{SEH12} S.Steentjes, D. Eggers and K. Hameyer, Application and verification of a dynamic vector-hysteresis model. \textit{IEEE Trans. Magnetics} \textbf{48} (2012) 3379-3382.
\bibitem{FLavHenr2013}  V. Francois-Lavet, F. Henrotte, L. Stainier, L. Noels and C. Geuzaine, An energy-based variational model of ferromagnetic hysteresis for finite element computations. \textit{J. Comp. Appl. Math.} \textbf{246} (2013) 243-250.
\bibitem{HSHG2014} F. Henrotte, S. Steentjes, K. Hameyer and C. Geuzaine, Iron loss calculation in steel laminations
at high frequencies. \textit{IEEE Trans. Magnetics} \textbf{50} (2014) 7008104.
\bibitem{B14} D. Lin, P. Zhou, A. Bergqvist, Improved vector play model and parameter identification for magnetic hysteresis materials. \textit{IEEE Trans. Magnetics} \textbf{50} (2014) 7008704.
\bibitem{Moreau} J. J. Moreau, Application of convex analysis to the treatment of elastoplastic systems. In: \textit{Lect. Notes in Math.} \textbf{503} (1976) 56-89 (Eds. P. Germain and B. Nayroles).
\bibitem{Torre} E. Della Torre, \textit{Magnetic hysteresis} (IEEE Press, Piscataway, 1999).
\bibitem{Appino} C. Appino, C. Ragusa, F. Fiorillo, Can rotational magnetization be theoretically assessed? \textit{Int. J. Appl. Electromagnetics and Mechanics} \textbf{44} (2014) 355-370.
\bibitem{LZC15} D. Lin, P. Zhou and N. Chen, A new vector hysteresis model based on series-distributed play hysterons. In: \textit{Compumag} 2015, Montreal, Canada.
\bibitem{ZVis} S. E. Zirka, Y. I. Moroz, Ph. Marketos and A. J. Moses, Viscosity-based magnetodynamic model of soft.
magnetic materials. \textit{IEEE Trans. Magnetics} \textbf{42} (2006) 2121-2132.
\bibitem{ZInv}  S. E. Zirka, Y. I. Moroz and R. G. Harrison, Inverse hysteresis models for transient simulation,
\textit{IEEE Trans. Power Delivery} \textbf{29} (2014) 552-559.
\bibitem{LPA} F. Liorzou, B. Phelps and D. L. Atherton, Macroscopic Models of Magnetization. \textit{IEEE Trans. Magnetics} \textbf{36} (2000) 418-428.
\bibitem{dARTV} M. d'Aquino, G. Rubinacci, A. Tamburrino and S. Ventre, Efficient numerical solution of magnetic field problems in presence of hysteretic media for nondestructive evaluation \textit{IEEE Trans. Magn.} \textbf{49}  (2013) 3167-3170.
\bibitem{Zloss} S. E. Zirka, Y. I. Moroz, Ph. Marketos and A. J. Moses, Loss separation in nonoriented electrical steels. \textit{IEEE Trans. Magnetics} \textbf{46} (2010) 286-289.
\bibitem{Zloss1} S. E. Zirka, Y. I. Moroz, Ph. Marketos and A. J. Moses, Viscosity-based magnetodynamic model of soft magnetic materials. \textit{IEEE Trans. Magnetics} \textbf{42} (2006) 2121-2132.
\bibitem{P97} L. Prigozhin, Analysis of critical-state problems
in type-II superconductivity. \textit{IEEE Trans. Supercond.} \textbf{7} (1997) 3866-3873.
\bibitem{Jackson} J. D. Jackson \textit{Classical Electrodynamics} (Wiley, NY, 1988).
\bibitem{bofm}To find $\vb_m=\vcurl a_m$, we note that
\begin{align*}
\vcurl[\nabla G(\vr-\vr')\times\vm(\vr',t)]=&\\
-\vm(\vr',t)(\Delta G)+\nabla G(\nabla\cdot\vm(\vr',t))-&\\(\nabla G\cdot\nabla)\vm(\vr',t)+
(\vm(\vr',t)\cdot\nabla)\nabla G=&\\ \vm(\vr',t)\delta(\vr-\vr')+\nabla[\nabla \cdot [G(\vr-\vr')\vm(\vr',t)]\,].
\end{align*}
Hence \begin{align*}\vcurl\int_{\Omega}\nabla G(\vr-\vr')\times\vm(\vr',t)\,d\vr'=&\\\vm(\vr,t)+\nabla\int_{\Omega}\nabla \cdot [G(\vr-\vr')\vm(\vr',t)]\,d\vr'.\end{align*}
\bibitem{BP14} J. W. Barrett and L. Prigozhin, Existence and approximation of a mixed formulation for thin film
magnetization problems in superconductivity. \textit{Math. Models and Meth. in Applied Sci.} \textbf{24} (2014) 991-1015.
\bibitem{Bess} L. A. Bessonov \textit{Theoretical Foundations of Electrical Engineering. Electromagnetic Field} [in Russian] (Uraite, Moscow, 2016).
\bibitem{Bur} J. Burkardt,  STRANG7 quadrature for triangles, see\\ \verb"http://people.sc.fsu.edu/~jburkardt"\\ \hspace{.5cm} \verb"/datasets/quadrature_rules_tri"
\bibitem{Kat} J. T. Katsikadelis \textit{Boundary Elements: Theory and Applications} (Elsevier, Amsterdam, 2002).

\end{thebibliography}
\end{document}